\documentclass[manuscript]{aastex}

\usepackage{natbib}
\usepackage{color}
\usepackage{dashrule}
\usepackage{pict2e}
\usepackage{tikz}

\shorttitle{Observations of a Pulse-driven Surge by SDO/AIA}
\shortauthors{Kayshap et al.}

\begin{document}


\title{Study of the Kinematics and Plasma Properties of A Solar Surge Triggered due to Chromospheric Activity in AR11271}


\author{P.~Kayshap\altaffilmark{1,2}}
\affil{Aryabhatta Research Institute of Observational Sciences (ARIES), Manora Peak, Nainital-263 129, India}

\author{Abhishek K. Srivastava\altaffilmark{1}}
\affil{Aryabhatta Research Institute of Observational Sciences (ARIES), Manora Peak, Nainital-263 129, India}

\author{K. Murawski\altaffilmark{2}}
\affil{Group of Astrophysics,UMCS, ul. Radziszewskiego 10, 20-031 Lublin, Poland}
\email{kmur@kft.umcs.lublin.pl}


\altaffiltext{1}{Aryabhatta Research Institute of Observational Sciences (ARIES), Manora Peak, Nainital-263 129, India.}
\altaffiltext{2}{Group of Astrophysics,UMCS, ul. Radziszewskiego 10, 20-031 Lublin, Poland.}
\begin{abstract}
We observe a solar surge in NOAA AR11271 using 
SDO AIA 304 \AA \ image data on 25 August, 2011. The surge rises 
vertically from its origin upto a height of $\approx$65 Mm with a
terminal velocity of $\approx$100 km s$^{-1}$, and thereafter 
falls and fades 
gradually. The total life time of the surge was $\approx$20 min. 
We also measure the temperature and density distribution 
of the observed surge during its maximum rise, and found an
average temperature and density of $2.0$ MK and $4.1\times$10$^{9}$ cm$^{-3}$, respectively.
The temperature map shows the expansion and mixing of cool
plasma lagging behind the hot coronal plasma along the surge.
As SDO/HMI temporal image data does not show any detectable evidence of the significant photospheric magnetic field 
cancellation for the formation of the observed  surge, we infer that 
it is 
probably driven by magnetic reconnection generated thermal
energy in the lower chromosphere. 
The radiance (thus mass density) oscillations near the base of the surge are also evident, which 
may be the most likely signature of its formation by a
reconnection generated 
pulse.
In support of the present observational base-line of the triggering of the surge
due to chromospheric heating, we devise a numerical model 
with conceivable implementation of
VAL-C atmosphere and a thermal pulse as an initial trigger. 
We find that the pulse
steepens into a slow shock at higher altitudes that
triggers plasma perturbations exhibiting the
observed features of the surge, e.g., terminal velocity, 
height, width, life-time, and heated fine structures near its base.
\end{abstract}




\section{Introduction}
Various types of plasma ejections are significant in the solar atmosphere 
to transport mass and energy at diverse spatio-temporal scales
\citep[e.g.,][and references cited there] {Boh75,Innes97,Kuj07,Kat07,Len07,Boris09,Sri11,Mor12}. 
The magnetically structured lower solar atmosphere is dynamically filled with different types of jet-like
phenomena at short spatial scales, e.g., spicules, mottles, fibrils, anemone jets,
tornadoes \citep[e.g.,][and references cited there]{Bar04,Has06,Bart07,
Bar11,Bohm12}. The large-scale plasma jets (polar jets, surges, sprays etc)
are also important to channel mass 
and energy into upper corona \citep[e.g.,][and references cited there]
{Sri11,Wu12}. The magnetic reconnection  and magnetohydrodynamic (MHD) wave activity 
are found to be the basic mechanisms for driving such plasma ejecta in the 
solar atmosphere \citep[e.g.,][and references cited there]
{Yoko95,Yoko96,Innes97,Bar04,Has06,Kuj07,Kat07,Cir07,Bart07,Bar07,Dip07,Par10,Boris09,Mur10,Kam10,Mur11,Mor12}. 

Among various types of solar jets, 
the solar surges are cool jets, typically formed by a plasma that is usually 
visible in H$_{\alpha}$ and other chromospheric and coronal lines \citep{Ster20}. The solar 
surges are most likely triggered by magnetic reconnection and they 
exhibit different episodes of heating and cooling \citep{Kur07}.
Solar surges are mostly associated with flaring regions
and places of transient and dynamical activities in the solar atmosphere. 
The solar surges and their evolution at various temperatures have been observed
extensively in association with magnetic field
and flaring activities of the solar atmosphere \citep[e.g.,][and references cited there]
{Sch94,Chae99,Yushi03,Liu04,Jia07,Wu12}. 
Inspite of the direct magnetic reconnection
and photospheric magnetic field emergence and cancellation \citep{Gai96,Ster20,Liu05,Wu12},
the solar surges can also be triggered by impulsive 
generation of pressure pulse \citep{Shib82, Ster93}, and may also 
be associated with the explosive events \citep{Maj09}.
However, understanding of their exact drivers
requires
further investigations, and more than one mechanism may work 
for the surge evolution depending
upon the local plasma and magnetic field conditions. Therefore, study 
of the driving mechanisms of any particular surge is always the forefront
of solar research that may also be useful probing of the localized 
conditions and magnetic activities in the solar atmosphere.

In the present paper, we describe empirical results derived from 
the observations of a surge that originated from the 
western boundary of an active region NOAA AR11271 on 25 August, 2011. We 
study the kinematics and plasma properties of 
this surge, as well as
its most likely trigger mechanism. We select a distinct
surge that was not related with the recurrent 
surge activities in AR11271, and apply a numerical model 
that reproduces its observed properties. Our model suggests
that the trigger of the observed surge may be due to the reconnection generated thermal pulse
in the lower solar atmosphere. 
In Sect. 2 we describe the observations of the surge, its kinematics, 
and derived plasma properties. We also investigate the magnetic fields at the base of the surge 
using Helioseismic Magnetic Imager (HMI)
data and discuss 
its consequences. We report on the numerical
model of the observed surge and present numerical results in Sect.~3
and conclude with the discussions. 

\section{STUDY OF THE KINEMATICS, PLASMA, AND MAGNETIC PROPERTIES OF A SOLAR SURGE TRIGGERED FROM AR11271}

For this study we have used the imaging and magnetic field observations from instruments onboard SDO. 
AIA is a multi-channel full-disk imager \citep{Lem11} with resolution elements of 0.6$"$ and a cadence of 
12 s. For this study we have used data of 171 \AA\ (coronal) and  304 \AA\ (upper chromospheric/TR) filters. 
HMI is designed to study the photospheric magnetic field and takes full-disk images at a cadence 
of 3.75 s to provide  Doppler intensity, magnetic field, and 
full vector magnetic field at a cadence of 45 s \citep{Sch12}.
SDO/AIA and SDO/HMI data have been obtained from Solar-Soft cut-out service established at Lokheed Martin Space Research Laboratory, USA 
$\footnote{http://www.lmsal.com/get\_aia\_data/
}$, which are corrected for flat-field and spikes. For further cleaning and calibration of this data, we also run aia\_prep subroutine of SSWIDL.

We use an image sequence of a solar surge in AR11271 on 25 August 2011 during 09:11--09:31 UT
as observed in the 304 \AA \hspace{0.05cm} and 193 \AA \hspace{0.05cm} filters, which represent 10$^5$ K and 10$^6$ K plasma, respectively.
The surge originated near the western
boundary of AR11271  as shown in Fig.~1.
These images are overlaid by SDO/HMI magnetic field contours
at a level of $\pm$800 G : yellow (blue) stays for 
positive (negative) polarity.
It is clearly
evident that the surge occurred near a positive polarity region (yellow) in the
north-west side of a big negative polarity spot (shown by blue circle). The surge plasma mostly emits 
in the 304 \AA\ filter sensitive to the plasma temperature around $\sim$10$^{5}$ K 
, but does not show much emission
in the coronal filter.
However, some emission in the hotter coronal channel is visible up to the height of 20 Mm from the foot point of the surge (cf.,
right panel of Fig.~1). 
Therefore, the surge consists of multi-temperature plasma.

\subsection{Kinematics of the Surge}

Fig.~2 represents the spatio-temporal evolution of the surge material using the sub-field 
data. It is clear that the surge is being
activated at 09:11 UT  near the western boundary of 
NOAA AR11271, and faded around 09:21 UT. 
The observed surge was not 
associated with the recurrent spray 
surge activity in AR11271 \citep{Kayshap12}. 
This image sequence shows that the surge reaches upto a projected 
height of $\approx$65 Mm in 
$\approx$660 s. The terminal speed of the surge is found to be $\approx$100 km s$^{-1}$. 
This surge is well collimated with a maximum width of $\approx$7 Mm and does not show
any significant lateral expansion.
Its down falling time is $\approx$500 s in which it also faded gradually.
Top-panel of Fig.~3 shows the distance-time measurement along the observed surge.
It is evident that the denser core material of the surge first rises up
and then decelerates. Some fainter and less denser plasma reaches much higher 
into the solar atmosphere, and falls freely thereafter.
The distance-time map across the surge at a particular height around $\approx$15 Mm
is shown in the bottom panel.
The radiance periodicity
is observed on 3.0 min time-scale, which may be the signature of the
periodic enhancement of mass density in the surge.
This may provide the 
clues about the periodic driver of the surge, which 
will be discussed in detail in Sec.~3. 
It is also possible that the periodicity of 3.0 min near the surge
base may be due to the swirling motion 
as recently observed in spectroscopic data \citep[e.g.,][and references cited therein]{Er12,Bart12}. 
However, such small-scale motion is below 
the detection limit in our data.
Moreover, the radiance oscillations occur higher in the transition region,
and therefore, in the present case it may be the most likely evidence of the periodic signature
of the arrival of slow shocks in the upper chromosphere/transition region.

\subsection{Estimations of the Average Temperature and Density of the Surge}

In the present section, we carried out emission measure 
and temperature distribution analysis, as well as the tracing and estimation of
density along the surge by extensively using the automated method developed by
\citet{Asc11}. Understanding such plasma properties
provide more clues about the driver of the surge, while comparing 
with the numerical results. 
SDO/AIA data provides information of the emission measure and estimation of the average 
density and temperature \citep{Asc11}.
We use
full-disk SDO/AIA in all EUV channels at the time of the maximum rise of the surge around 
09:22 UT on 25 August, 2011. We calibrate and clean the data using aia\_prep subroutine 
of SSW IDL as well as co-align the AIA images 
as observed in various AIA filters by using the co-alignment test as described by
\cite{Asc11}.

Using the automated code developed by \cite{Asc11}, 
we obtain emission measure and temperature maps for six AIA filter full-disk and co-aligned
images (304 \AA, 171 \AA, 193 \AA, 94 \AA, 335 \AA, 211 \AA ) 
in the temperature between 0.5-9.0 MK. 
The emission measure map in the 
top-left panel of Fig.~4 shows that the surge has significant emissions 
from its denser core part. The emissions from its leading edge is comparatively low. 
Top-right panel of Fig.~4 shows the distribution 
of temperature in and around the surge location. It is evident that heating is
confined near its foot-point. Greenish-yellow regions represent the highest temperature in and
around the footpoint of the surge.
The reddish region shows the plasma at the base of the surge with a temperature of 6.2-6.3 MK. 
Above the heated area near the base of the surge, the cool plasma is maintained at 
sub-coronal temperatures as represented by blue-black colors. It is also mixed with the 
comparatively hot plasma. The leading edge of the surge is, however, maintained 
at comparatively high coronal temperatures. 
The heating near the base of the surge probably drives it,
and the cool plasma moves behind the hot plasma at its 
leading edge. It is also mixed with the hot plasma, and constitutes 
the multi-temperature surge.

Using the automated procedure \citep{Asc11}, we trace the loop segments 
and surge 
in the co-spatial and co-temporal images of various wavelengths of SDO/AIA data. The input free
parameters for this automated tracing of the surge is summarized in Table 1.
The bottom panel of Fig.~4 shows that the surge
material is erupting along open field lines, indicated by white arrow, from the western part of active region boundary. 
The surge consists of multi-temperature plasma. The contribution of cool surge plasma (T=0.1 MK) as evident in 304 \AA\ channel, is represented by yellow colour.
 For the estimation of the plasma 
properties (temperature and density) along the surge material and its overlying 
diffuse corona, we trace the larger segment of this coronal structure using another set of free parameters 
(cf., Table 1) and implement the best forward fit of the possible DEM 
solutions. 
We choose
this set of free parameters on the basis of trial and error, and it gives the best DEM
forward fit in the case of 171 \AA\ fluxes at each pixel of the considered region around the surge
and its overlying atmosphere.
In Fig.~5, the estimated density and temperature are displayed. The chi-square of the estimation 
in the forward modeling is good-enough up to 70 Mm along the surge as well as 
upto 140 Mm in its overlying diffuse corona. 
The average temperature and density of the surge material are respectively
2.0 MK and 4.1$\times$10$^{9}$ cm$^{-3}$. We can estimate these physical 
parameters in all the traced structures in all AIA channels. However, we 
display the best fit results of the estimation of the density and temperature 
along the surge and its overlying corona in 171\AA\  channel as shown in Fig.~5. 

\subsection{SDO/HMI Investigations of the Magnetic Field Properties at the Base of the Surge}

The surge occurred 
at the north-western side of the  big sunspot with negative polarity. It was 
located at the outer layer of a
positive polarity region (cf., snapshot on 09:00 UT in Fig.~6). 
The surge is 
shown as intensity contour (blue) of co-aligned AIA data on HMI snapshot of 09:21 UT in Fig.~6. 
The morphology of the origin site of the observed surge is almost similar to that observed recently 
by \cite{Wu12} in association with the recurrent surges from NOAA 10884 on 25 October 2003.
 However, we do not observe any strong signature of the magnetic field annihilation as well as moving magnetic features (MMFs) in our present 
observational data as previously observed by \cite{Wu12}. 
Therefore, we infer that the observed surge may not be triggered by  strong photospheric magnetic field activities. It is evident
that tiny negative magnetic polarities have been evolved near the surge location, as well as the
major positive polarity region has shown some areal enhancement towards the western direction. 
Therefore, such rearrangements of the magnetic
polarities around the surge region may produce more complexity in the overlying magnetic 
flux-tubes \citep{Roy73, Can96}, e.g., in low-lying loops and an arch crossing near the base of the surge (cf., 
09:11 UT AIA 304 \AA\ snapshot of Fig.~2; HMI-AIA snapshot on 09:20 UT in Fig.~6). 
Therefore, the chromospheric 
reconnection between the open field lines as well as surrounding 
closed fields may trigger the surge.
The localized brightening and heating at the same location is evident where dark arch type flux-tube and low-lying
closed loops are crossing near the
surge base (cf., 09:11 UT snapshot of Fig.~2, and right panel of Fig.~4), which may generate the 
thermal pulse probably triggering the observed plasma perturbations. 
In the next section, in support of the present observational base-line of the triggering of the surge
due to chromospheric heating, we devise a numerical model 
with implementation of
VAL-C atmosphere and a thermal pulse as an initial trigger,
which re-produce the observations closely. 

\section{A NUMERICAL MODEL FOR THE PULSE-DRIVEN SOLAR SURGE}\label{SECT:NUM_MODEL}
To reproduce the observed surge, our model system assumes a
gravitationally-stratified solar atmosphere 
that is described by
the ideal two-dimensional (2D) 
MHD equations:
\begin{equation}
\label{eq:MHD_rho}
{{\partial \varrho}\over {\partial t}}+\nabla \cdot (\varrho{\bf V})=0\, ,
\end{equation}
\\
\begin{equation}
\label{eq:MHD_V}
\varrho{{\partial {\bf V}}\over {\partial t}}+ \varrho\left ({\bf V}\cdot \nabla\right ){\bf V} =
-\nabla p+ \frac{1}{\mu}(\nabla\times{\bf B})\times{\bf B} +\varrho{\bf g}\, ,
\end{equation}
\\
\begin{equation}
\label{eq:MHD_p}
{\partial p\over \partial t} + \nabla\cdot (p{\bf V}) = (1-\gamma)p \nabla \cdot {\bf V}\, ,
\end{equation}
\\
\begin{equation}
\label{eq:MHD_B}
{{\partial {\bf B}}\over {\partial t}}= \nabla \times ({\bf V}\times{\bf B})\, , 
\hspace{3mm}
\nabla\cdot{\bf B} = 0\, .
\end{equation}
Here ${\varrho}$, ${\bf V}$, ${\bf B}$, $p = \frac{k_{\rm B}}{m} \varrho T$, $T$, $\gamma=5/3$,
${\bf g}=(0,-g)$ with its value $g=274$ m s$^{-2}$, $m$, $k_{\rm B}$, are respectively 
the mass density, flow velocity, magnetic field, gas pressure, temperature,
adiabatic index, gravitational acceleration, mean particle mass, and Boltzmann's constant.
It should be noted that we do not invoke the radiative cooling and 
thermal conduction in our model for simplicity reasons since we simulate the cool
surge ejecta in the solar atmosphere.
\vspace{-1.30cm}
\subsection {Equilibrium configuration}
%
%
%
We assume that at its equilibrium the solar atmosphere is still (${\bf V}_{\rm e}={\bf 0}$) with a force-free magnetic field,
\begin{equation}\label{eq:B_e}
(\nabla\times{\bf B}_{\rm e})\times{\bf B}_{\rm e} = {\bf 0}\ , 
\end{equation}
such that it satisfies a current-free condition,
%
$\nabla \times {\bf B}_{\rm e}={\bf 0}$, and it is specified by the magnetic flux function, $A$,
as
\begin{equation}\label{eq:B_e1}
{\bf B}_{\rm e}=\nabla \times (A\hat {\bf z})\, .
\end{equation}
%
Here the subscript '${\rm e}$' corresponds to equilibrium quantities.
We set an arcade magnetic field by choosing
%
\begin{equation}
A(x,y) = B_{\rm 0}{\Lambda}_{\rm B}\cos{(x/{\Lambda}_{\rm B})} {\rm exp}[-(y-y_{\rm r})/{\Lambda}_{\rm B}]\, 
\end{equation}
%
%
%
%
with 
$B_{\rm 0}$ 
being
the magnetic field at $y=y_{\rm r}$, and the magnetic scale-height is
\begin{equation}
{\Lambda}_{\rm B}=\frac{2L}{\pi}\, .
\end{equation}
%
We set and hold fixed $L=100$ Mm. For such choice the magnetic field vectors are weakly curved (Fig.~\ref{fig:T_B_lines}, bottom panel). 

As a result of Eq.~(\ref{eq:B_e}) 
the pressure gradient is balanced by the gravity force,
\begin{equation}
\label{eq:p}
-\nabla p_{\rm e} + \varrho_{\rm e} {\bf g} = {\bf 0}\, .
\end{equation}
%
%
%
With the ideal gas law and the $y$-component of 
Eq.~(\ref{eq:p}), we 
arrive at 
\begin{equation}
\label{eq:pres}
p_{\rm e}(y)=p_{\rm 0}~{\rm exp}\left[ -\int_{y_{\rm r}}^{y}\frac{dy^{'}}{\Lambda (y^{'})} \right]\, ,\hspace{3mm}
\label{eq:eq_rho}
\varrho_{\rm e} (y)=\frac{p_{\rm e}(y)}{g \Lambda(y)}\, ,
\end{equation}
where
\begin{equation}
\Lambda(y) = \frac{k_{\rm B} T_{\rm e}(y)}{mg}
\end{equation}
is the pressure scale-height, and $p_{\rm 0}$ denotes the gas 
pressure at the reference level that we choose in the solar corona at $y_{\rm r}=10$ Mm.

We take
an equilibrium temperature profile $T_{\rm e}(y)$ for the solar atmosphere
derived from the VAL-C atmospheric model of \citet{Ver81}. 
Having specified $T_{\rm e}(y)$ (see Fig.~7, top panel) with Eq.~(\ref{eq:pres})
we obtain the corresponding gas pressure and mass density profiles.
 
The transition region (TR) is located at $y\simeq 2.7$ Mm. 
Above the TR we assume an extended solar corona.
Below the solar chromosphere, the temperature minimum 
is located at $y\simeq 0.9$ Mm.

%
%

%
%
\vspace{-1.30cm}
\subsection{Perturbations}
%
We initially perturb
the equilibrium impulsively by a Gaussian 
pulse in 
a gas pressure 
viz.,
\begin{equation}
p(x,y,t=0) = p_{\rm e}(y) + \nonumber \\
p_{\rm e}(y)A_{\rm p} 
\exp\left[ 
-\frac{(x-x_{\rm 0})^2} {w_{\rm x}^2}
-\frac{(y-y_{\rm 0})^2} {w_{\rm y}^2} 
\right]
\, .
\label{eq:perturb}
\end{equation}
Here $A_{\rm p}$ is the amplitude of the pulse, $(x_{\rm 0},y_{\rm 0})$ is its initial position and
$w_{\rm x}$, $w_{\rm y}$ denote its widths along the $x$- and $y$-directions, respectively. 
We set and hold fixed $A_{\rm p}=80$, 
$x_{\rm 0}=0$ Mm, $y_{\rm 0}=1.75$ Mm, 
$w_{\rm x}=2.25$ Mm, and $w_{\rm y}=0.2$ Mm. 
%
%
%
\vspace{-1.20cm}
\subsection{Results of the Numerical Simulations}
Equations (\ref{eq:MHD_rho})-(\ref{eq:MHD_B}) are solved numerically using the FLASH code
\citep{Lee09,FLASH99}. This code uses a second-order unsplit Godunov solver 
with various slope 
limiters and Riemann solvers, as well as adaptive mesh refinement (AMR).
We set the simulation box 
of $(-12.5,12.5)\, \times (0,125)\,$ Mm$^{2}$ along the $x$- and $y$-directions 
(Fig.~\ref{fig:T_and_blocks}, bottom panel) 
and set and hold fixed all plasma quantities at all boundaries of the simulation region to their equilibrium values 
which are given by Eqs.~(\ref{eq:B_e1}) and (\ref{eq:eq_rho}). 
These fixed boundary conditions performed much better than transparent boundaries, 
leading only to negligibly small numerical 
reflections of wave signals from these boundaries. 

In our modeling, we use an AMR grid with a minimum (maximum) level of 
refinement set to $4$ ($7$) (Fig~\ref{fig:T_and_blocks}). The refinement strategy is based on 
controlling numerical errors in 
mass density, which results in an excellent resolution of steep spatial profiles and
greatly reduces numerical diffusion at these locations.

Fig.~9 displays the spatial profiles of the plasma temperature (color maps) 
and velocity (arrows) 
resulting from the initial pulse of Eq.~(\ref{eq:perturb}), which splits into counter-propagating pulse trains. The downward propagating
part is reflected from the dense ambient plasma of the photosphere. This reflected part lags behind the originally upward propagating 
pulses, which become slow shocks. Because the plasma is initially pushed upward, the under-pressure results in the region below the initial pulse. This under-pressure sucks up 
comparatively cold chromospheric plasma, which lags behind the shock front at higher temperature. 
As a result, the pressure gradient force works against gravity and forces 
the chromospheric material to arrive into the solar corona in form of a rarefaction wave. 
At $t=200$ s this shock reaches the altitude of $y\simeq 65$ Mm and 
the 
cold plasma edge 
is
located at $y\simeq 35$ Mm, which is below the slow shock. 
The reason for the material being lifted up is 
the rarefaction of the plasma behind the shock front, which leads to low pressure there.
The next snapshot (top middle panel) is drawn for $t=400$ s. At this time 
the chromospheric plasma is located below at $y\simeq 55$ Mm. 
At the next moment, $t=600$ s (top right panel) 
the chromospheric plasma shows its developed phase, reaching the level of $y=65$ Mm, 
which approximately matches the observational data of Fig.~2 
(top right panel).
The cool surge slows down while propagating upward. At $t=800$ s it arrives at the level of $y\simeq 70$ Mm 
and subsequently 
the surge already 
subsides
being 
attracted by gravity (bottom panels).
At the subsiding stage the heated fine structures develop as a result of interaction between the down-falling plasma.
During the evolution of the surge, the comparatively hot plasma is evident near the base of the simulated surge that 
qualitatively matches the observations. The leading front of the slow shock moving above the cool surge 
plasma may also leave some traces in the overlying diffused quiet-Sun atmosphere evident in the coronal filters 
(cf., right panel of Fig.1). However, the less denser and diffused overlying quiet-Sun may not be visible
as a tracer of a slow shock front.
The temporal evolution of the density reveals 
that the surge is denser near its base, while the density decreases
with the height (Fig.~10). This is consistent with the observational data.
\section{DISCUSSIONS AND CONCLUSION}\label{SECT:DISS}
In the present paper, we study the kinematics and plasma properties of the solar surge triggered in AR11271 on 25 August, 2011.
We also performed a 2 D numerical simulation using VAL-C model of the solar atmosphere as initial condition and the FLASH code \citep{FLASH99}, which approximately
reproduce the dynamics of the observed cool surge.
The surge rises 
vertically from its origin site up to a height $\approx$65 Mm with a
terminal velocity of $\approx$100 km s$^{-1}$, and thereafter 
falls and fades
gradually within its total life time of $\approx$20 min. 
We also measure the temperature and density distribution 
of the observed surge during its maximum rise, and its 
average temperature and density are 
respectively $2.0$ MK and $4.1\times$10$^{9}$ cm$^{-3}$.
We do not get any strong evidence of the photospheric magnetic field 
activity (i.e., MMFs, flux-emergence, field annihilation, etc) 
in support of the formation of the observed surge. The surge is
 most likely found to be driven by magnetic reconnection generated thermal
energy in the chromosphere. 

The numerical simulations exhibit a general scenario of rising and subsequently 
down-falling/fading of the observed surge with a number of 
observational features such as its height and width, ejection, approximate rising time-scale, and average 
rising velocity that are close to the observational data. The heated fine structures at the base 
of the surge are also matching well both in the observations and theory. The temperature distribution 
(Fig.~4) shows that the significant amount of cool plasma expands along the surge, while 
its leading edge is maintained at coronal temperatures. This scenario matches qualitatively 
with our simulation results when the comparatively cooler core plasma of surge lags 
behind the high temperature slow shock. The kinematical estimations of the surge (Fig.~3)
also reveals the acceleration of the surge material and its subsequent free fall  
due to gravity thereafter, which also matches well the simulation results.
However, some parts of the comparison between the numerical 
and observational data are approximate and only qualitative. This mismatch may result from 
simplified profiles of the parameters in our numerical model. 
Moreover, the real surge was excited in more complex plasma and magnetic field conditions at the 
boundary of AR11271. 

It is reported in literature that if the height of the energy release site 
is less (greater) than a critical height then the jet will be driven by 
crest shocks 
(directly by the pressure gradient force 
generated by explosion) \citep{Shib82}. However, this minimum critical height in the solar atmosphere 
depends upon the strength of pressure enhancement in time at the energy release site.
For the larger pressure enhancement, this height 
will be lower and vice versa. Under the conditions when the energy release site is 
higher in the solar atmosphere, 
the surge material may be accelerated
by the ${\bf j}{\times}{\bf B}$ force as a result of reconnection 
\citep{Yoko95, Yoko96} as observed by \cite{Nisu08}. Therefore, 
the surge dynamics may not be so simple like a thermal pulse driven jet,
and magnetic forces may also play some role in its formation. 
If the reconnection and thus the energy release site occurs in the
lower chromosphere, any ${\bf j}{\times}{\bf B}$ force driven flow may
generate slow mode MHD wave/shock which eventually accelerates
the jet in the upper chromosphere and transition region
\citep{Shib82}. We consider the energy release site and the triggering 
by the thermal pulse at $y$=1.75 Mm above the photosphere, and its amplitude A$_{p}$=80 times larger
than the ambient gas pressure. 
In our case, A$_{p}$ is beyond the parameters considered
by \cite{Shib82}. We aim to perform 
some parametric studies in order to compare numerical results with the findings of 
\cite{Shib82}, which will be devoted to our potential future 
studies. 
However, it is noteworthy that the numerical model we devised, is an extension of the 1D model of 
\cite{Shib82}. We adopted the temperature profile and a curved magnetic field in the MHD regime, therefore
our 2D model represents a natural extension of the hydrodynamic model of \cite{Shib82} in simulating 
the cool jet. 

%
%
\section{Acknowledgments}
We acknowledge the constructive  suggestions of the reviewer, which considerably improve our manuscript,
use of automated DEM and temperature analysis of solar structures 
developed by Prof. Markus Aschwanden, LMSAL, USA, and the use of the SDO/AIA observations for this study. 
The FLASH code has been developed by the DOE-supported ASC/Alliance
Center for Astrophysical Thermonuclear Flashes at the University of
Chicago.
This work has been supported by 
a Marie Curie International Research Staff Exchange Scheme Fellowship within the 7th
European Community Framework Program (K.M.). 
PK acknowledges UMCS, Lublin, Poland for the visit grant in 2011 where the part
of the work has been executed.
AKS acknowledges Shobhna Srivastava for patient encouragement during the work.
AKS also thanks Prof. K. Shibata for his valuable suggestions during 
the discussion on the simulation of solar jets.
KM thanks Kamil Murawski for his assistance in drawing numerical data. 
\bibliographystyle{apj} 
\bibliography{reference} 

\clearpage



\clearpage
\begin{deluxetable}{ccrrrrrrrrcrl}
\tabletypesize{\scriptsize}
\rotate
\tablecaption{The Summary of Input Free Parameters}
\tablewidth{0pt}
\tablehead{
\colhead{} & \colhead{Half} & \colhead{Threshold level} & \colhead{Min filling} & \colhead{Output} & \colhead{Minimum} & \colhead{Max traced} & \colhead{Min curvature} \\ 
\colhead{} & \colhead{Width} & \colhead{in standard} & \colhead{factor} & \colhead{resolution} & \colhead{length} & \colhead{structures} & \colhead{radius} \\
\colhead{} & \colhead{} & \colhead{flux deviation} & \colhead{} & \colhead{} & \colhead{} & \colhead{} & \colhead{}
}
\startdata
To trace the surge & 2 pixels & 1.0 & 0.35 & 5.0 & 0.01 & 10,000 & 25 \\
To estimate the plasma properties & 10 pixels & 1.0 & 0.15 & 5.0 & 0.04 & 10,000 & 60 \\
\enddata
\end{deluxetable}

\clearpage

\begin{figure*}
\centering
\mbox{
\includegraphics[scale=0.73, angle=90]{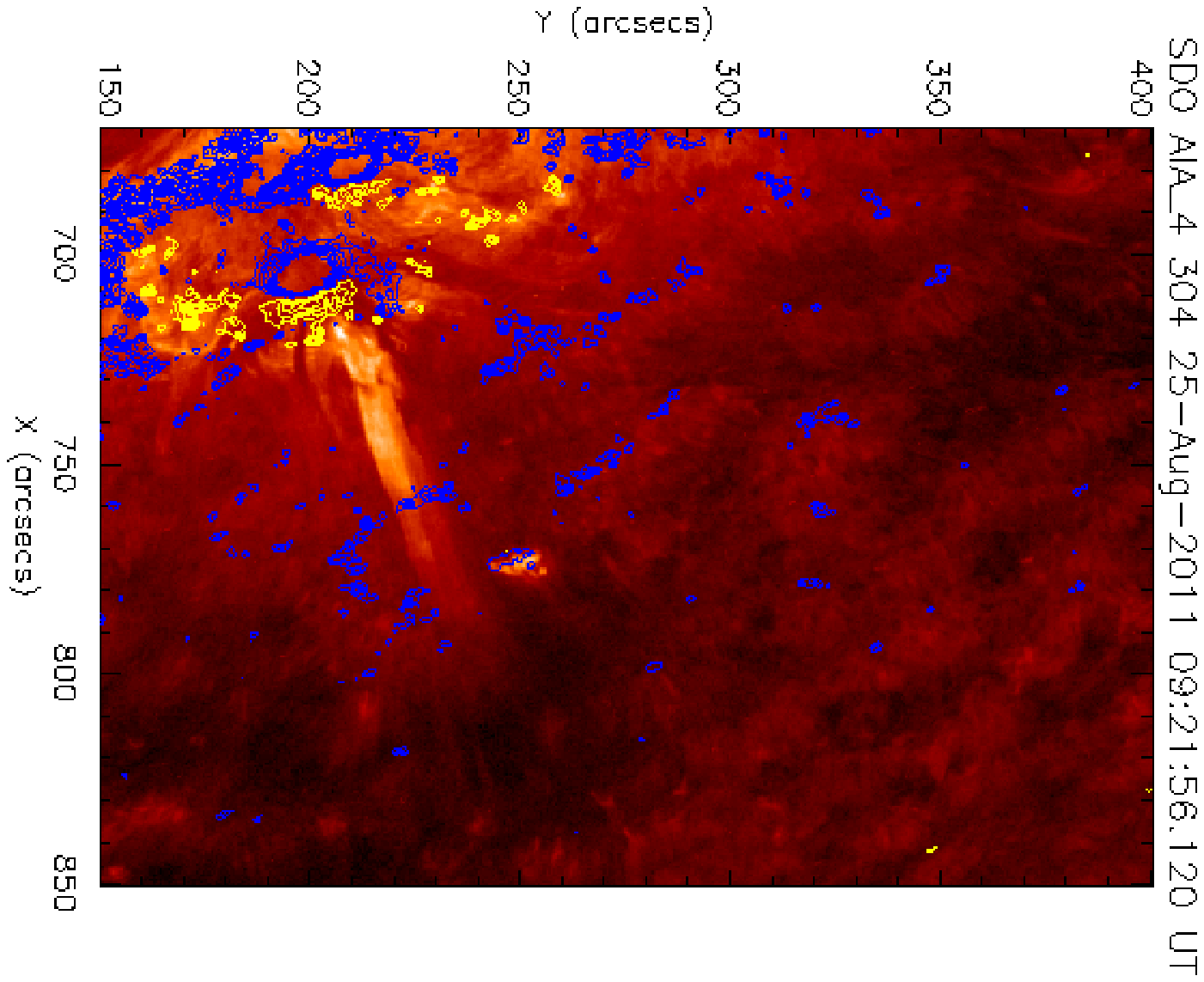}
\includegraphics[scale=0.73, angle=90]{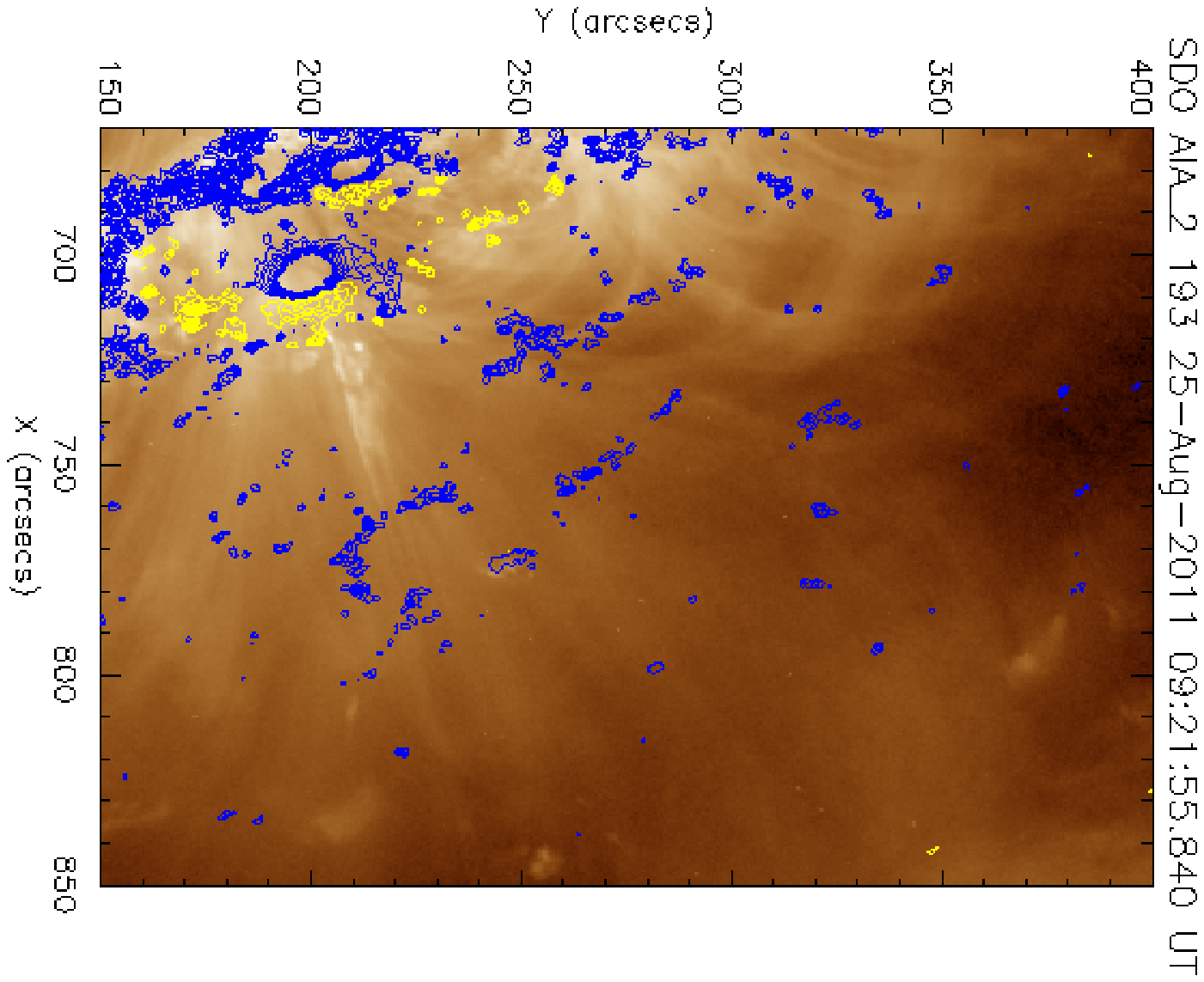}}
\caption{\small
SDO/AIA 193 \AA\ (right, yellow-brown map for 1.0 MK plasma) and 304 \AA\ (left, red-temperature map for 0.1 MK plasma) EUV images overlaid by
HMI magnetic field contours of maximum level $\pm$800 G. Yellow (blue) contours show
positive (negative) polarity of the magnetic field.
The surge origin is accompanied by heating above 
its foot-point. 
The observed surge is clearly evident in He II 304 \AA\ filter
that is sensitive to the low temperature plasma of 0.1 MK. 
Some heated part of the surge especially near its base up-to
the height of 20 Mm, is also
evident in the coronal filters (e.g., 193 \AA\ ) that
are sensitive to the plasma temperature around 1.0 MK. The surge is originated from the western boundary 
of the positive polarity situated in the North-West direction of the big sunspot with negative polarity 
as represented by the blue circular contour at X = 700$"$, Y = 200$"$.}
\label{fig:JET-PULSE}
\end{figure*}
%
\begin{figure*}
\centering
\mbox{
\hspace{0.1cm}
\includegraphics[scale=0.39, angle=90]{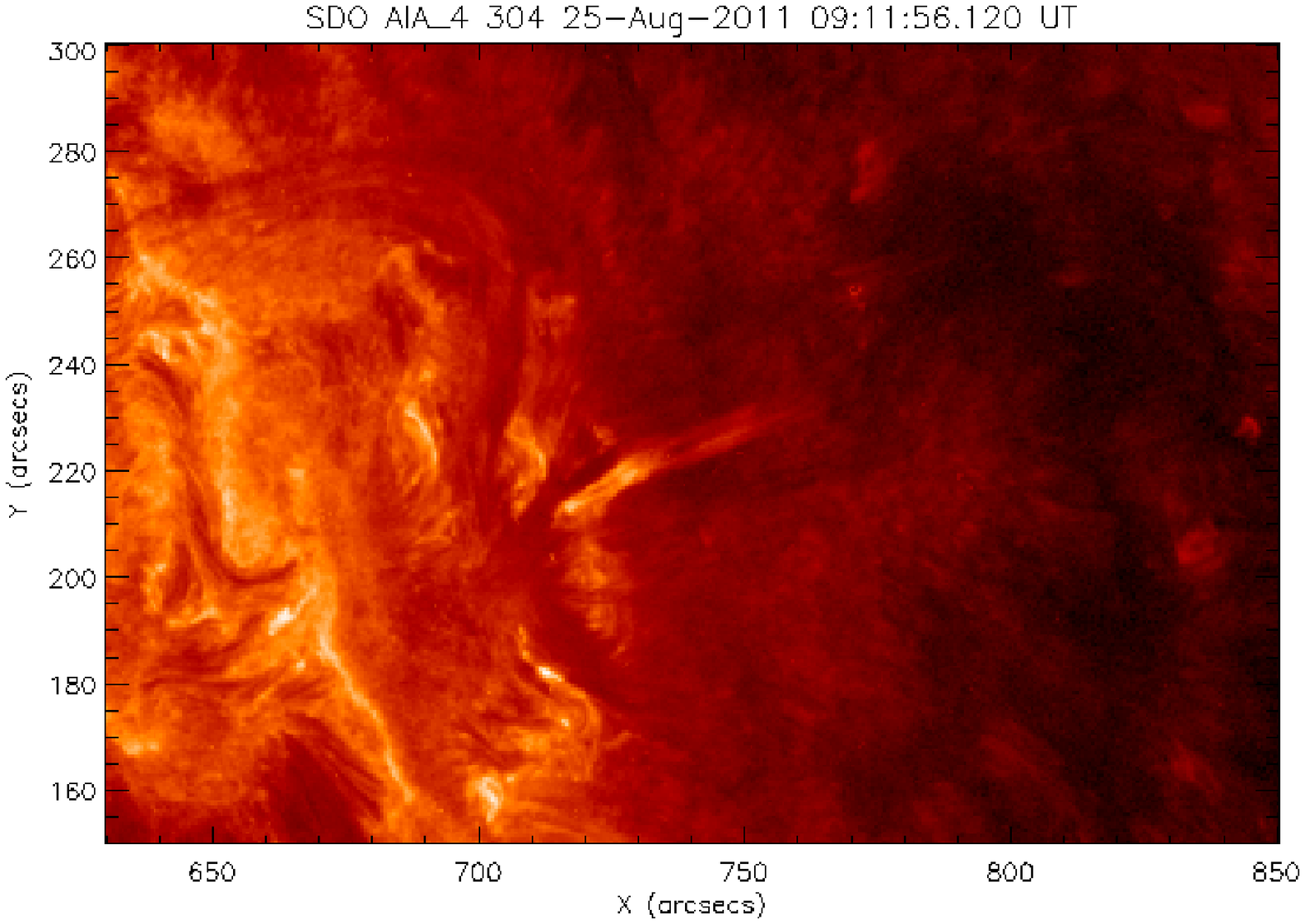}
$ \color{white} \put(-116,118){\vector(-1,-1){20}}\put(-155,128){Low Lying Loop Arches}$
\includegraphics[scale=0.39, angle=90]{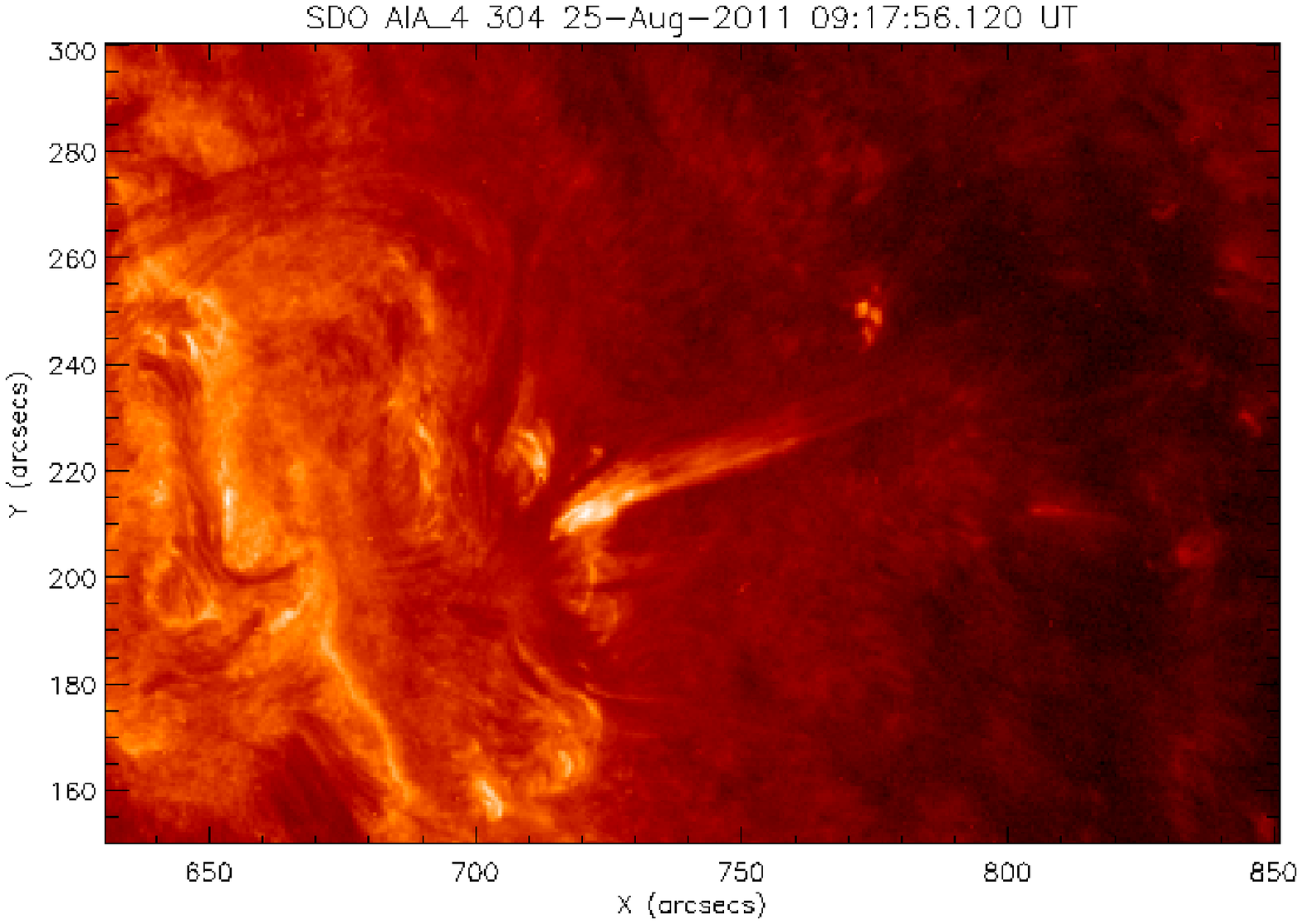}
}
\mbox{
\hspace{-0.2cm}
\includegraphics[scale=0.54, angle=0]{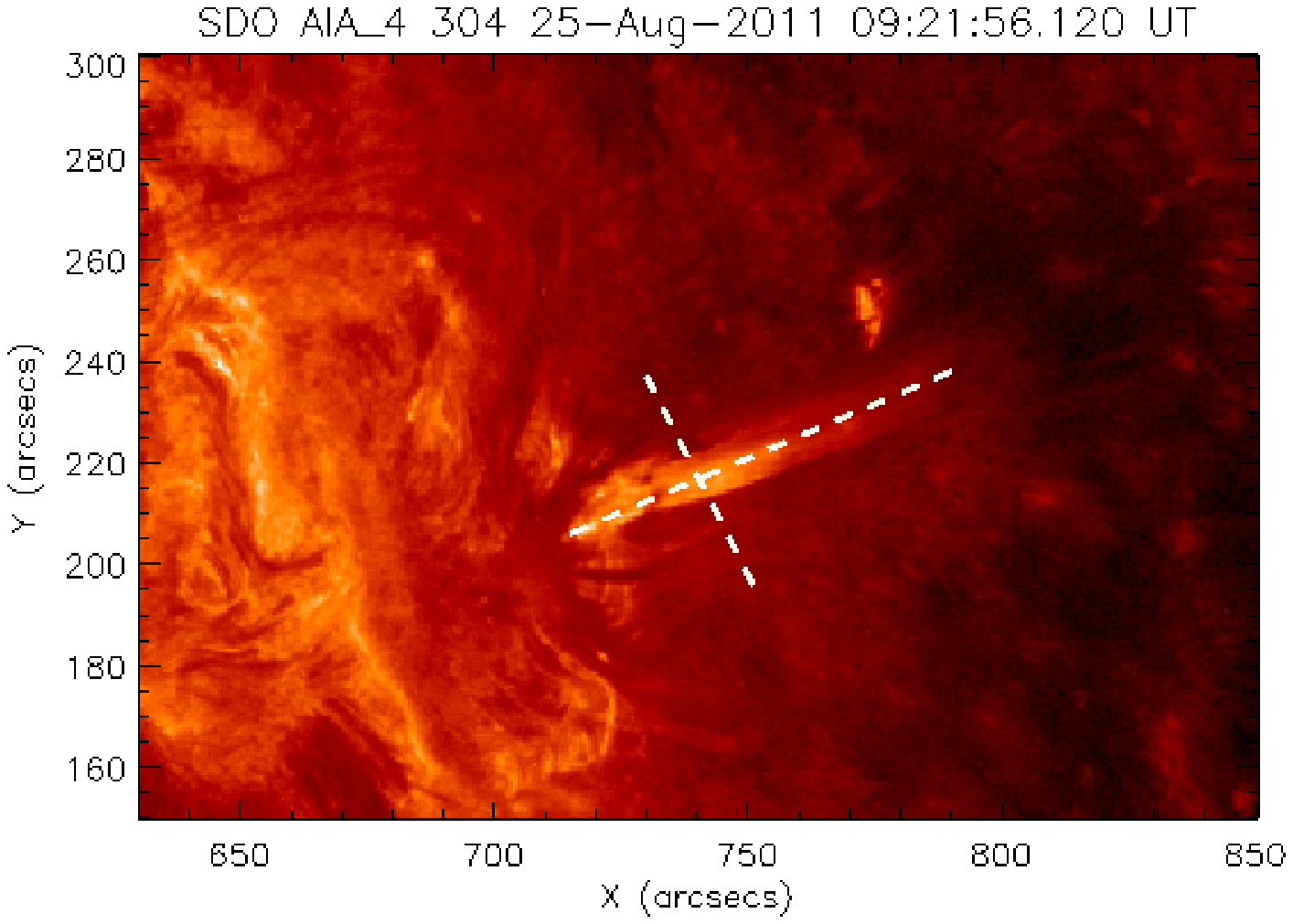}
\includegraphics[scale=0.39, angle=90]{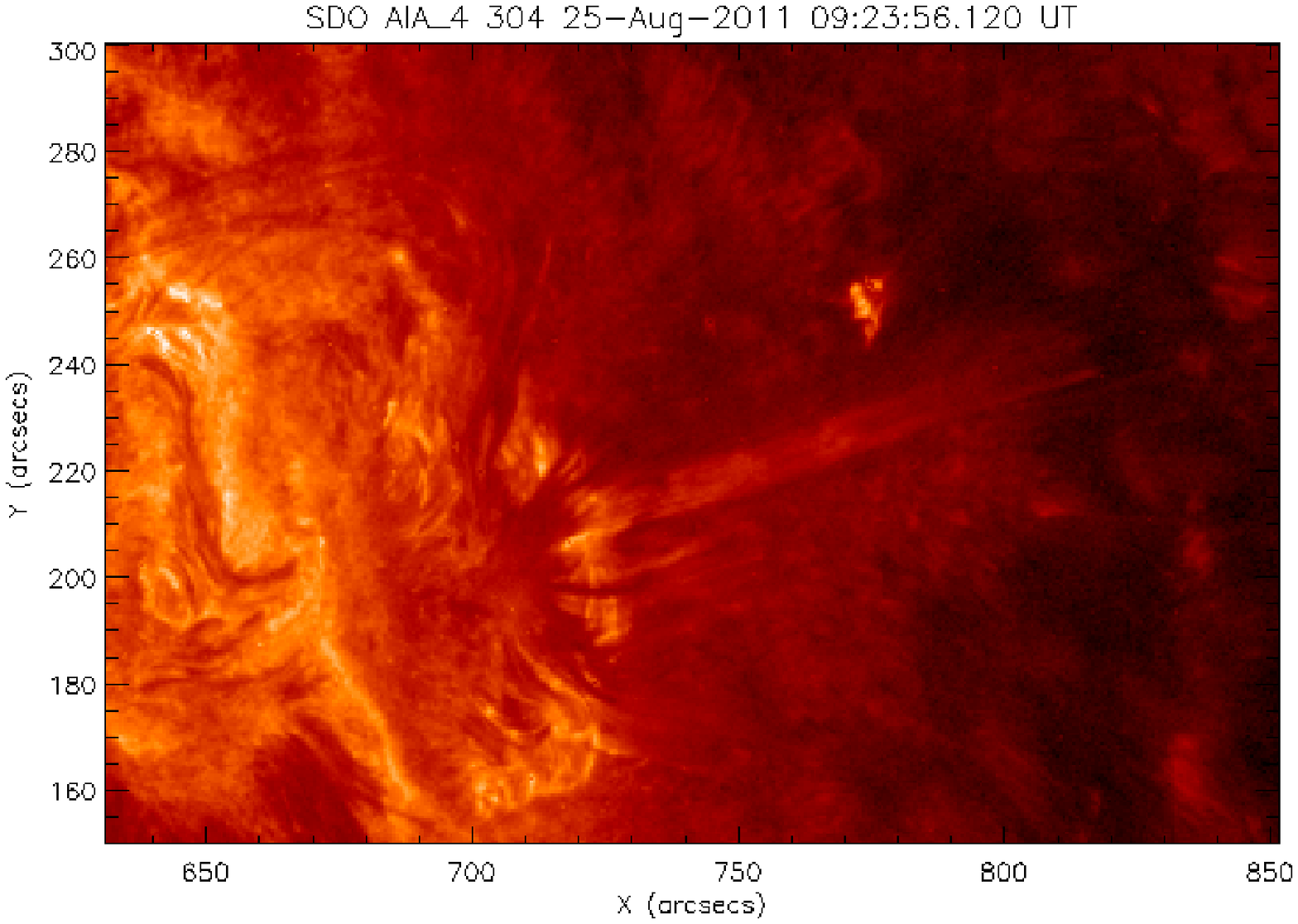}}
\hspace{-0.2cm}
\mbox{
\includegraphics[scale=0.39, angle=90]{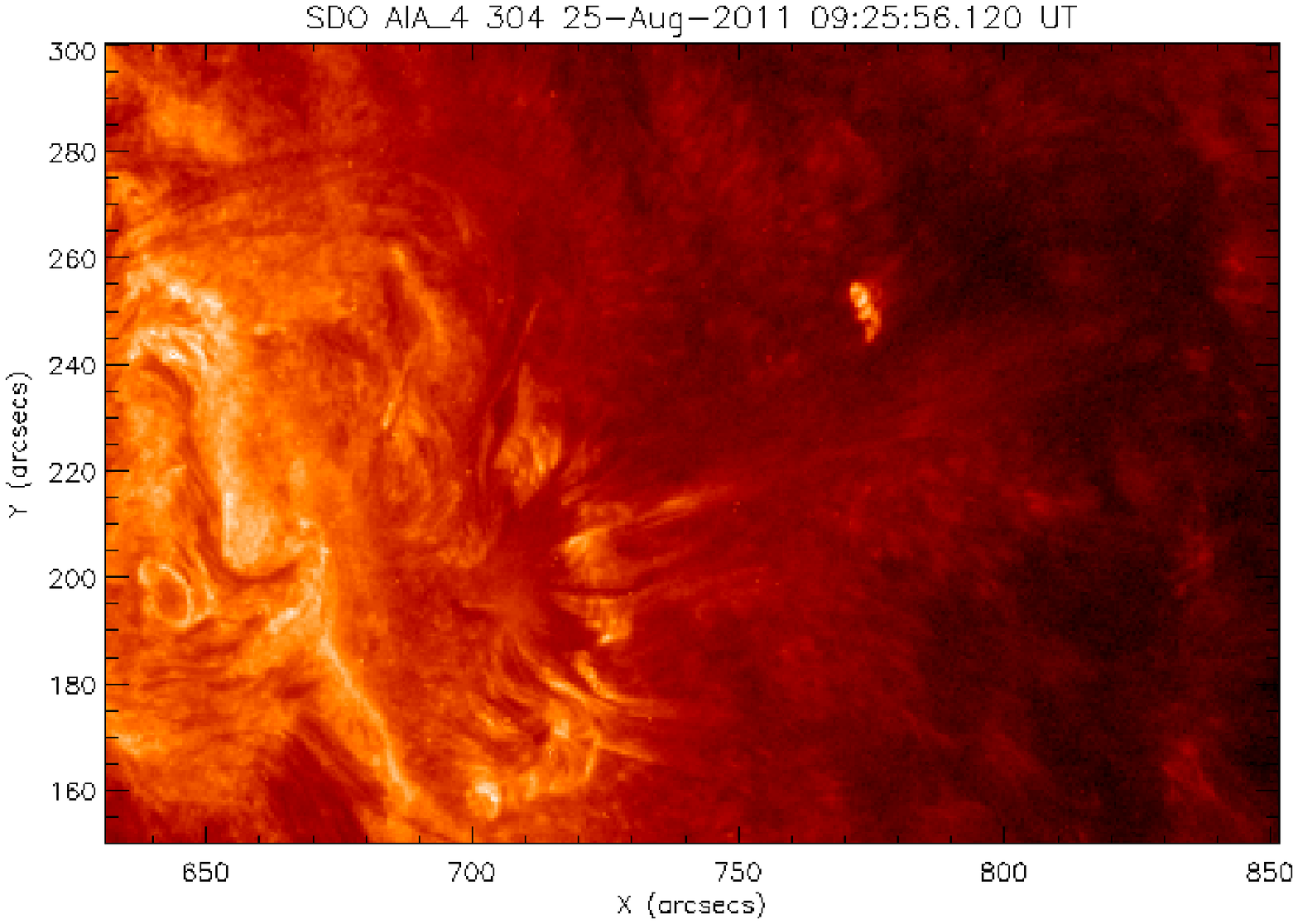}
\includegraphics[scale=0.39, angle=90]{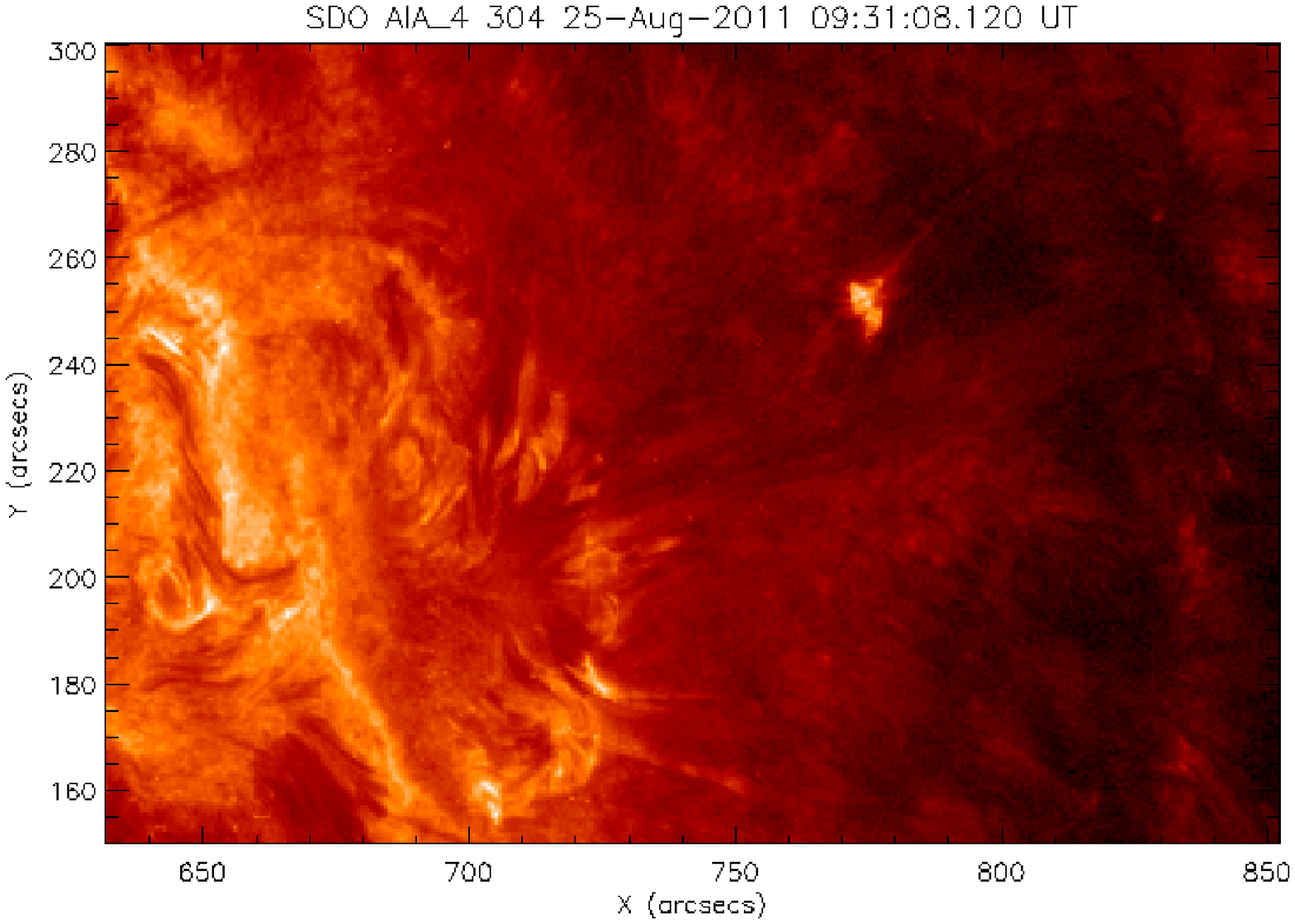}}
\caption{\small
SDO/AIA 304 \AA\ time sequence of a surge propagation higher into the corona from the boundary 
of an active region NOAA AR11271. The surge fades gradually with time, and the
material tries to settle down due to gravity. Horizontal and vertical
dotted lines on the snapshot of 09:21 UT show the two paths along which the Distance-Time measurements have been made. 
}
\label{fig:JET-PULSE}
\end{figure*}
\clearpage
\begin{figure*}
\centering
\includegraphics[scale=1.0, angle=0]{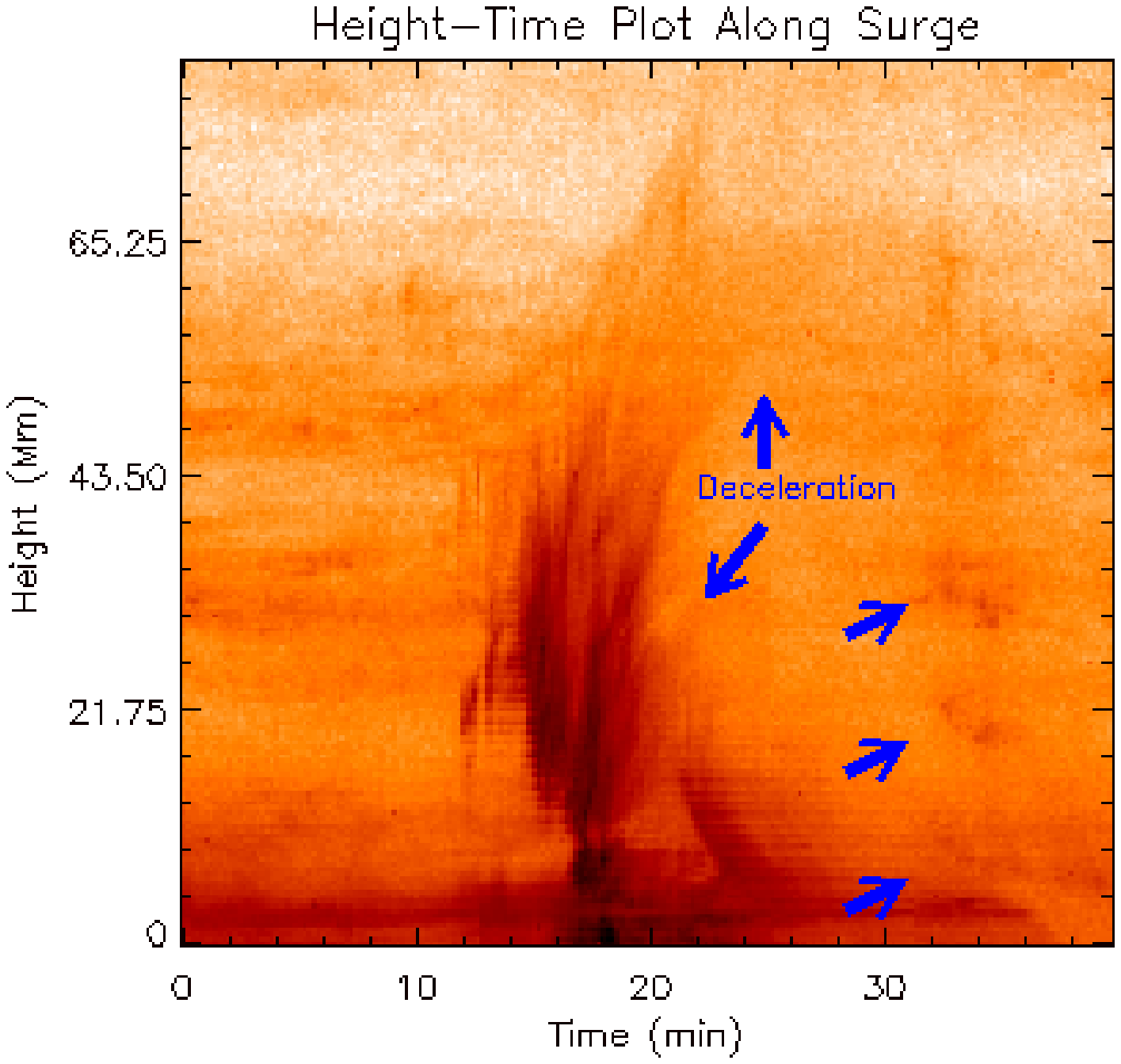}
\includegraphics[scale=0.7, angle=0]{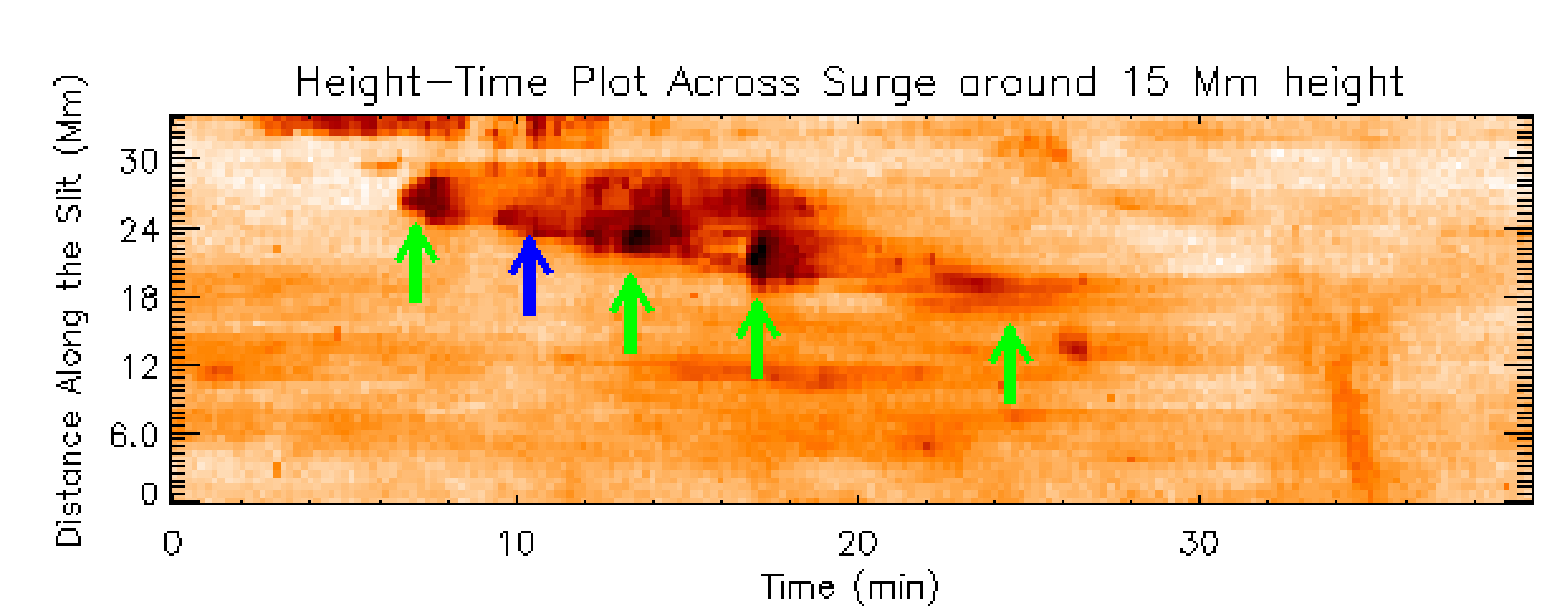}
\caption{\small The height-time plots : (i) Along the solar surge using He II
304 \AA\ data (top) where the distance is measured along the slit chosen along the surge;
 (ii) Across the surge at a height of $\approx$15 Mm from its base (bottom) where the distance is measured along the slit chosen across the 
surge. The positions of both the slits are shown over the 09:21 304 \AA\ snapshot in Fig.~2. 
}
\label{fig:JET-PULSE}
\end{figure*}
%
\clearpage

\begin{figure*}
\centering
\mbox{
\includegraphics[scale=0.50, angle=0]{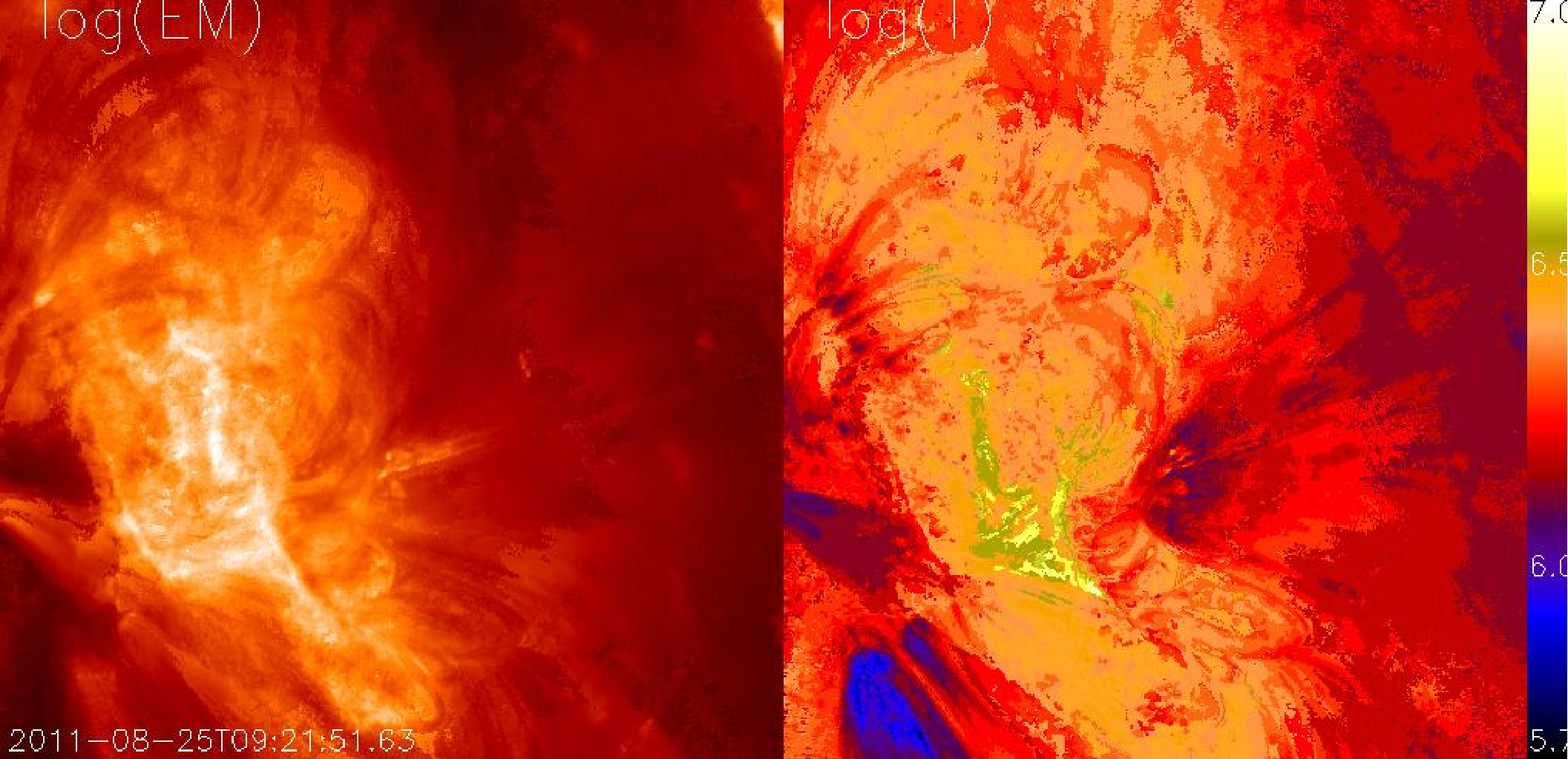}
$ \color{white} \put(-140,130){\vector(0,-1){40}} \put(-175,133){Heated Base}$
$ \color{white} \put(-128,43){\vector(0,1){50}} \put(-155,37){Cool Material}$
$ \color{white} \put(-110,83){\vector(0,1){20}} \put(-125,70){Hot Edge }$
$ \color{white} \put(-380,75){\vector(0,1){20}} \put(-400,60){Solar Surge }$
}
\mbox{
\includegraphics[scale=0.53, angle=0]{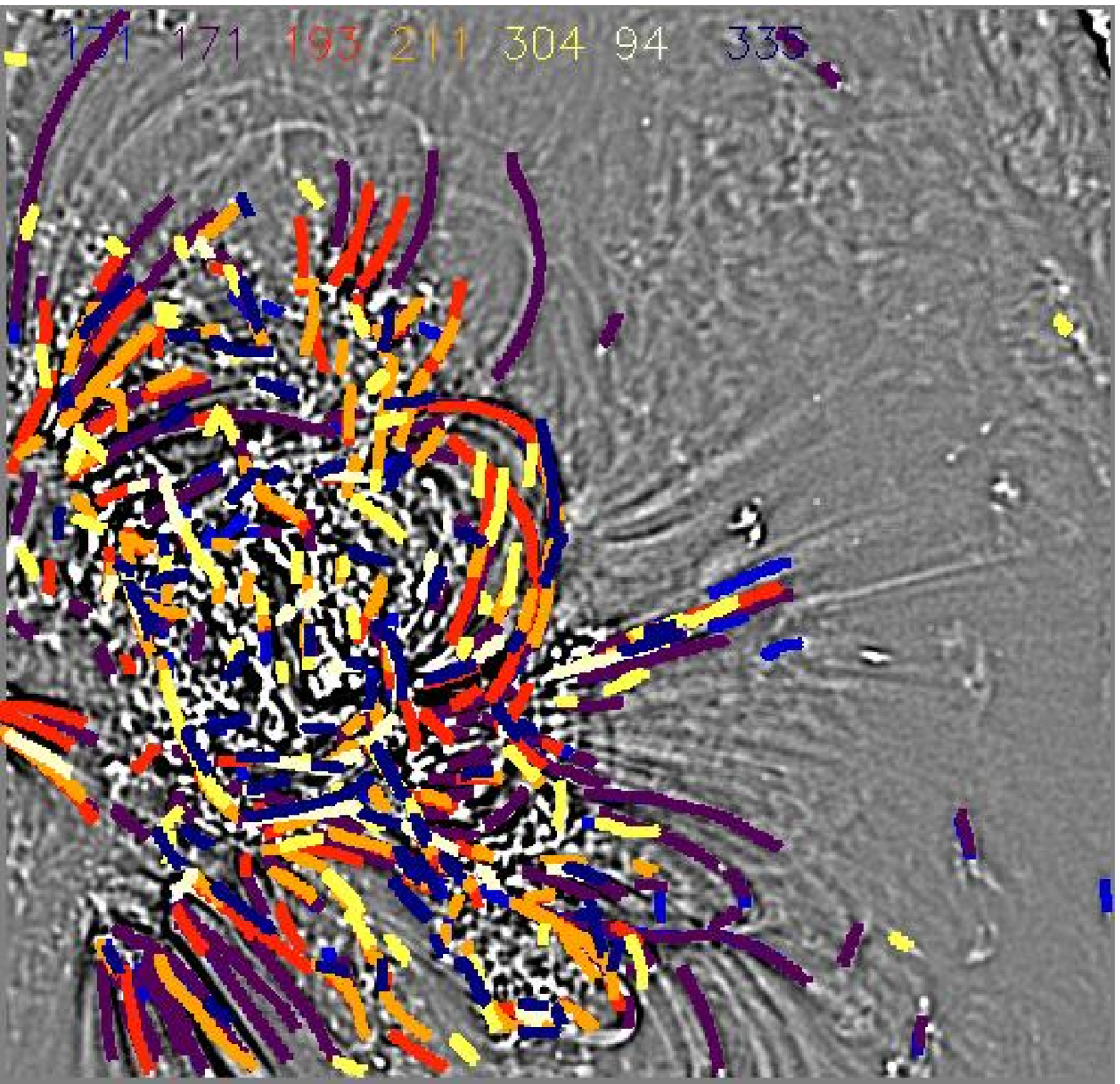}
\thicklines
$ \color{white} \put(-88,152){\vector(0,-1){15}} \put(-110,155){Solar Surge}$
}
\caption{\small The Emission Measure (EM) map (top-left) and 
temperature map (top-right panel) : Evidence of the multi-temperature
plasma forming the surge. The lower part of the surge consists of cool 
plasma, while the leading edge exhibits coronal
temperatures. Bottom panel shows the automated 
tracing of the plasma threads as visible at various temperatures 
along the open field lines of the surge as shown by arrow using the methodology of \cite{Asc11}.
Colour codes for the emitting plasma in different AIA channels are shown at the top of this snapshot.}
\label{fig:JET-PULSE}
\end{figure*}
%

\clearpage
\begin{figure*}
\centering
\includegraphics[scale=0.80, angle=0]{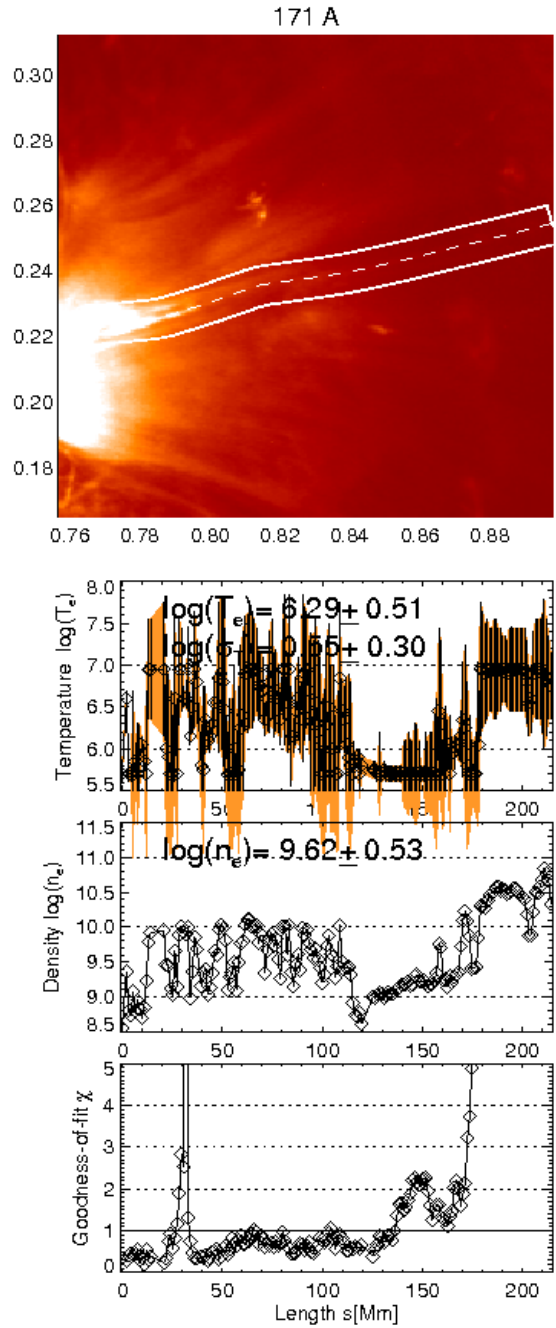}
\caption{\small The AIA 171 \AA\ image (top-panel) displays the sub-field containing the surge
and its overlying diffuse corona. 
The image axes are scaled as a fraction of the solar radius. 
This panel shows the 171 \AA\ image in which 
the box shows the area along the surge and its overlying diffuse corona along which the physical parameters
are evaluated.
The estimated plasma properties,
e.g., the integrated density, temperature along the 
surge (dashed-line) during its maximum rise are displayed in two consecutives middle-panels.} 
\label{fig:JET-DIAG}
\end{figure*}
%
\begin{figure*}
\centering
\includegraphics[scale=0.96, angle=0]{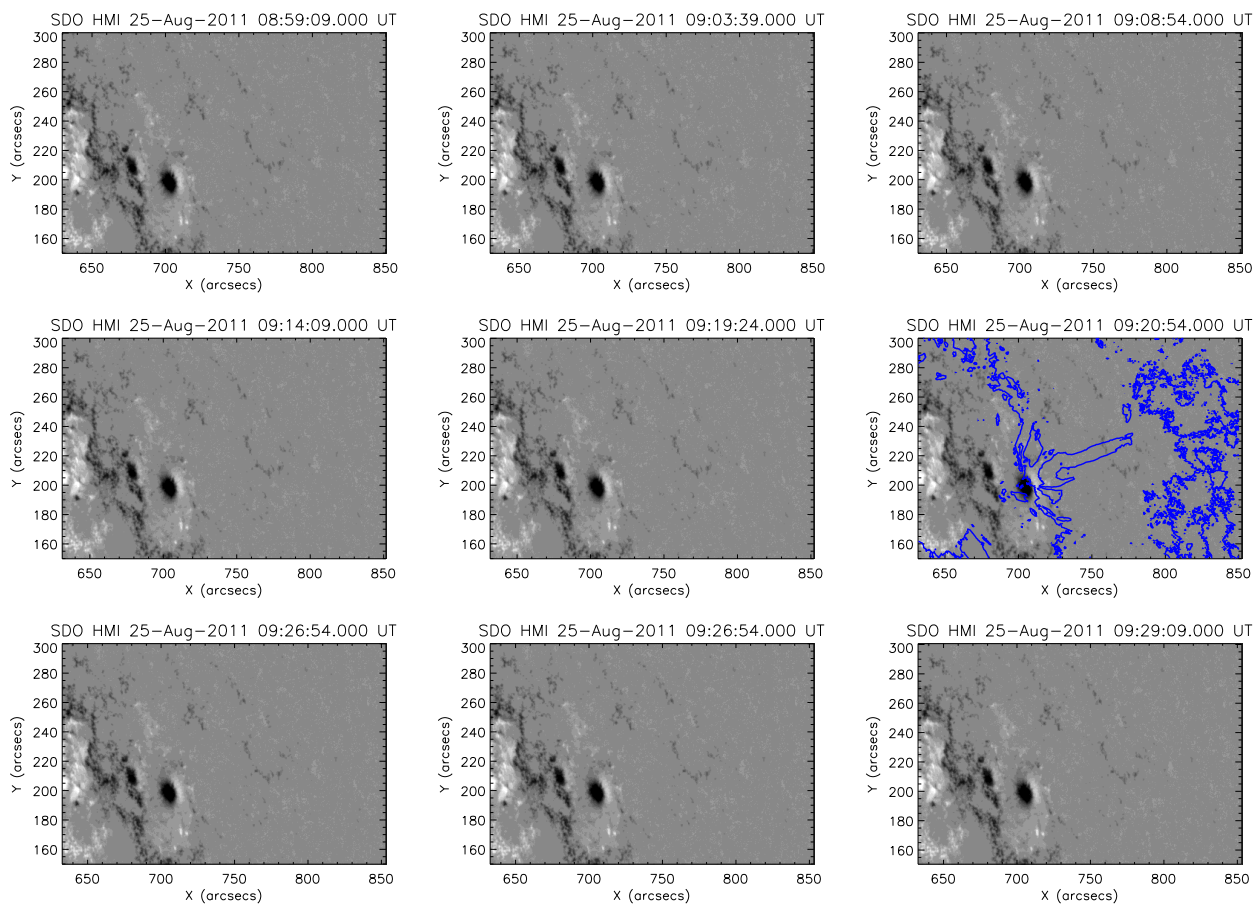}
\caption{\small
Co-aligned SDO/HMI time sequence of the photospheric magnetic field 
near the origin of the surge. 
The blue surge is 
shown as intensity contour on HMI snapshot of 09:21 UT. 
The blue loop arches on both the sides of the surge may cause possible reconnection 
at its base to heat and subsequently launch a thermal pulse for its acceleration.
}
\label{fig:JET-PULSE}
\end{figure*}

\clearpage
\begin{figure*}
\centering
\includegraphics[scale=0.60, angle=0]{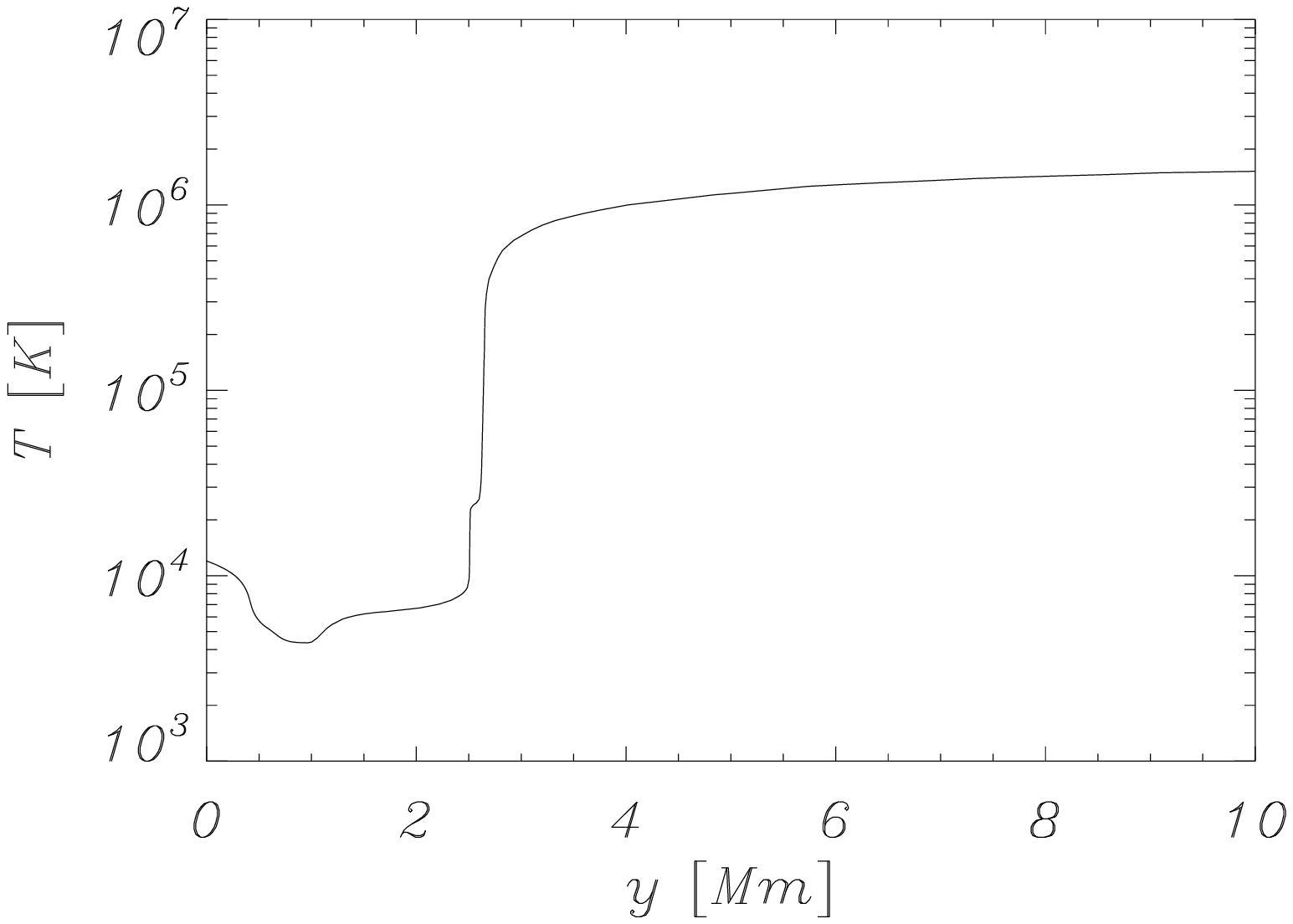}
\includegraphics[scale=0.60, angle=0]{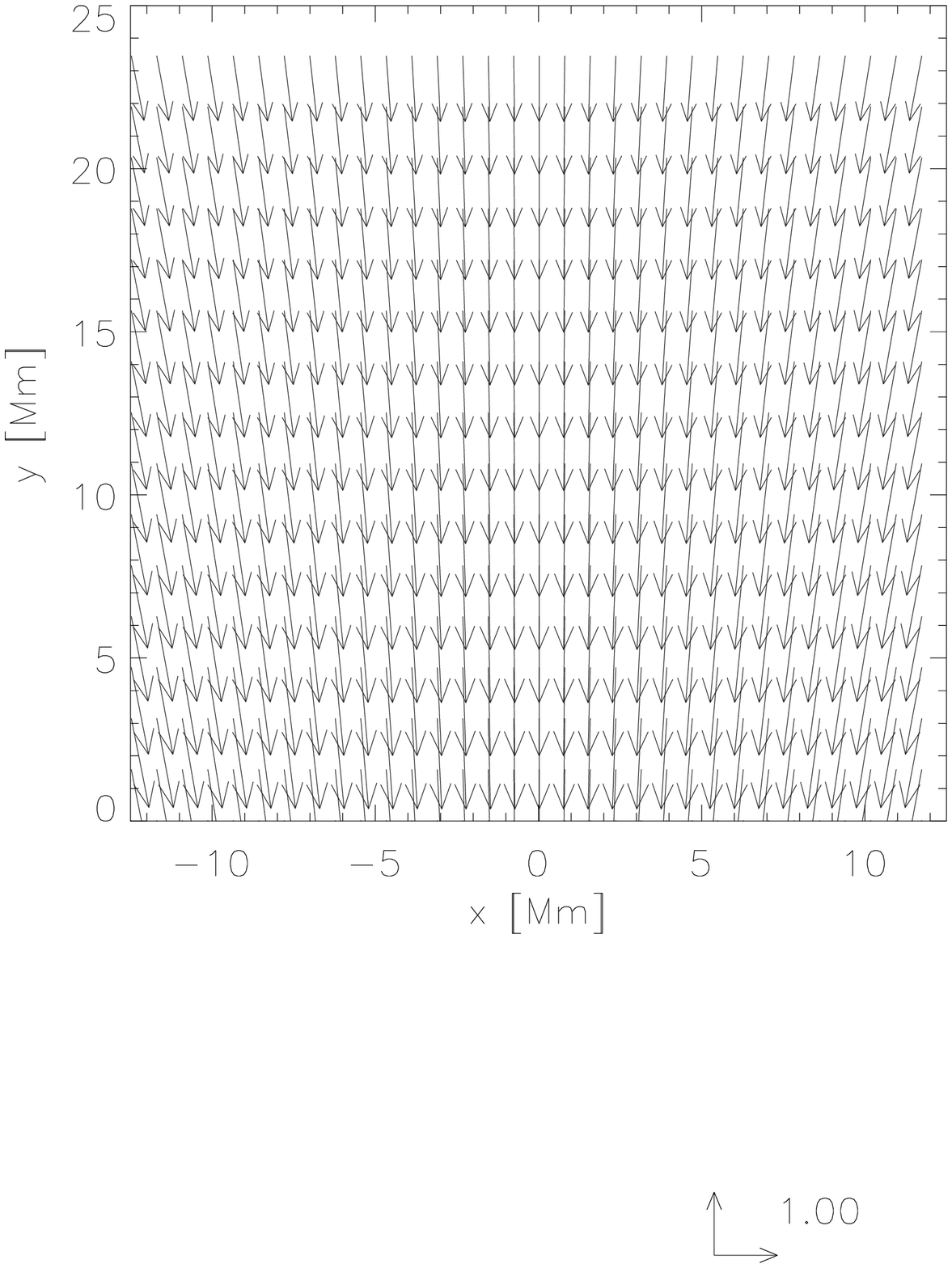}
\caption{\small 
Equilibrium profile of solar temperature (top panel) and magnetic field vectors (bottom panel).
The magnetic field is expressed in units of 44 gauss and a part of the simulation region is displayed.
}
\label{fig:T_B_lines}
\end{figure*}

\clearpage
\begin{figure*}
\centering
\includegraphics[scale=0.80, angle=0]{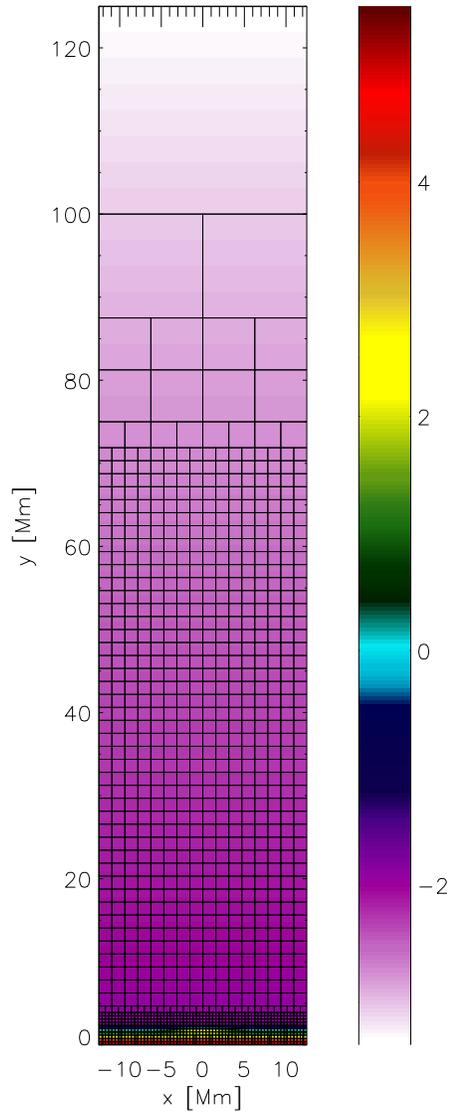}
\caption{\small 
The grid system with block boundaries represented by
solid lines as adopted in the simulation. The initial pressure perturbation of the
system is displayed by the horizontally elongated patch at a location $(0 Mm, 1.75 Mm)$.
}
\label{fig:T_and_blocks}
\end{figure*}

\clearpage

%
\begin{figure*}

\centering
\mbox{
  \includegraphics[scale=0.55]{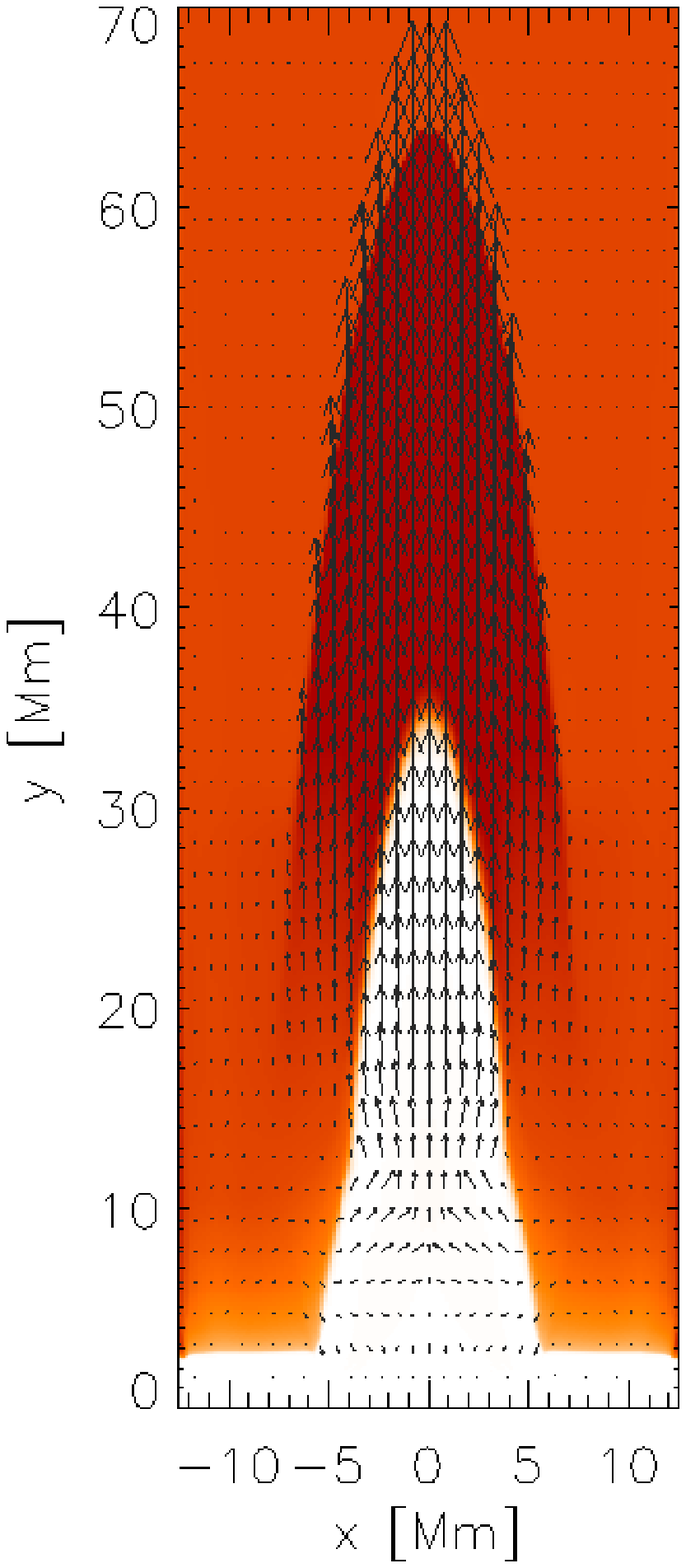}
  \includegraphics[scale=0.55]{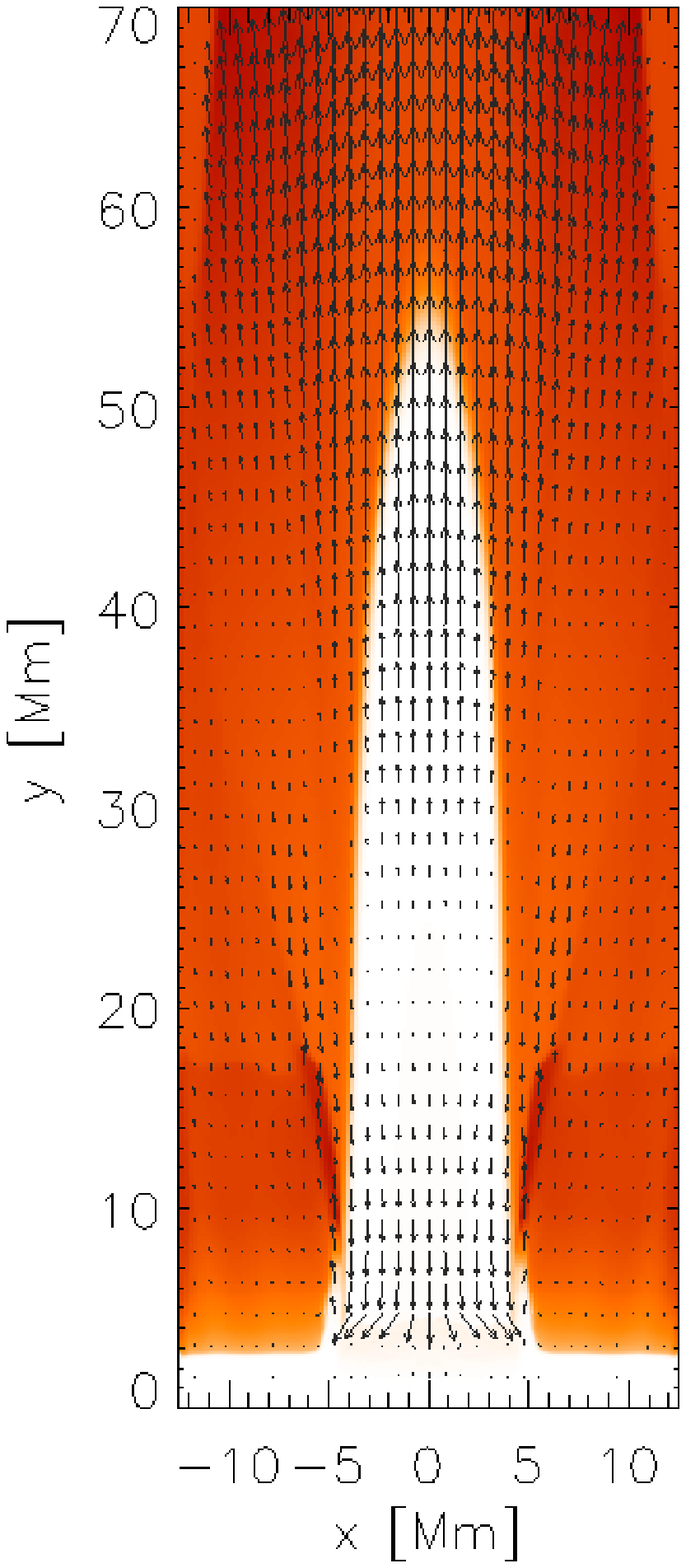}
  \includegraphics[scale=0.55]{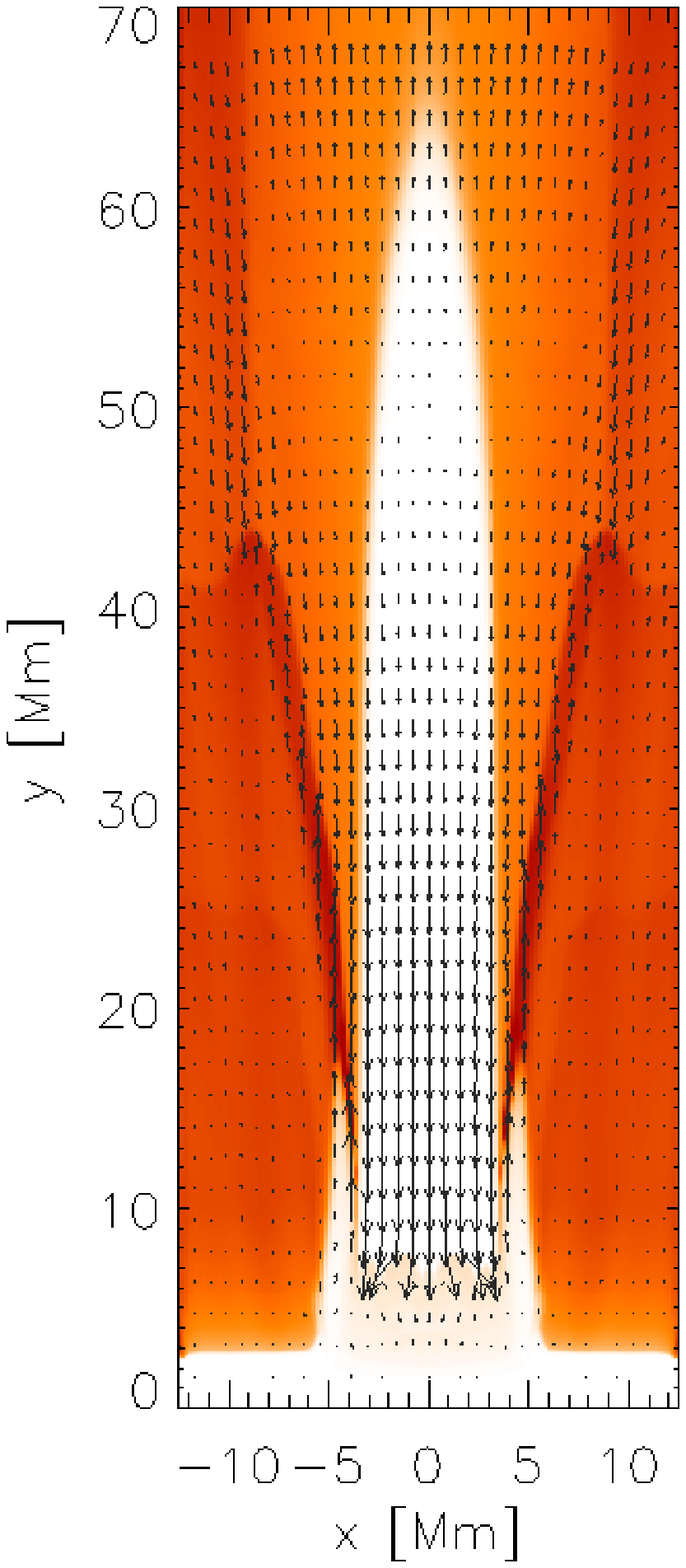}
  \includegraphics[scale=0.55]{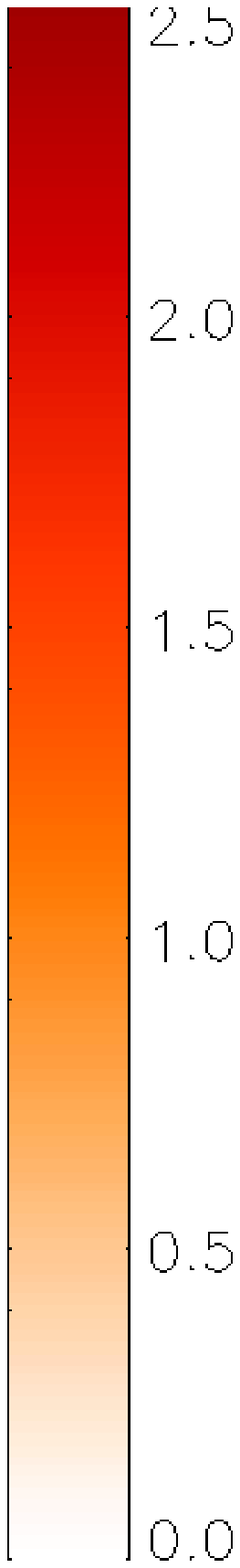}
}
\mbox{
  \hspace{-1.8cm}
  \includegraphics[scale=0.55]{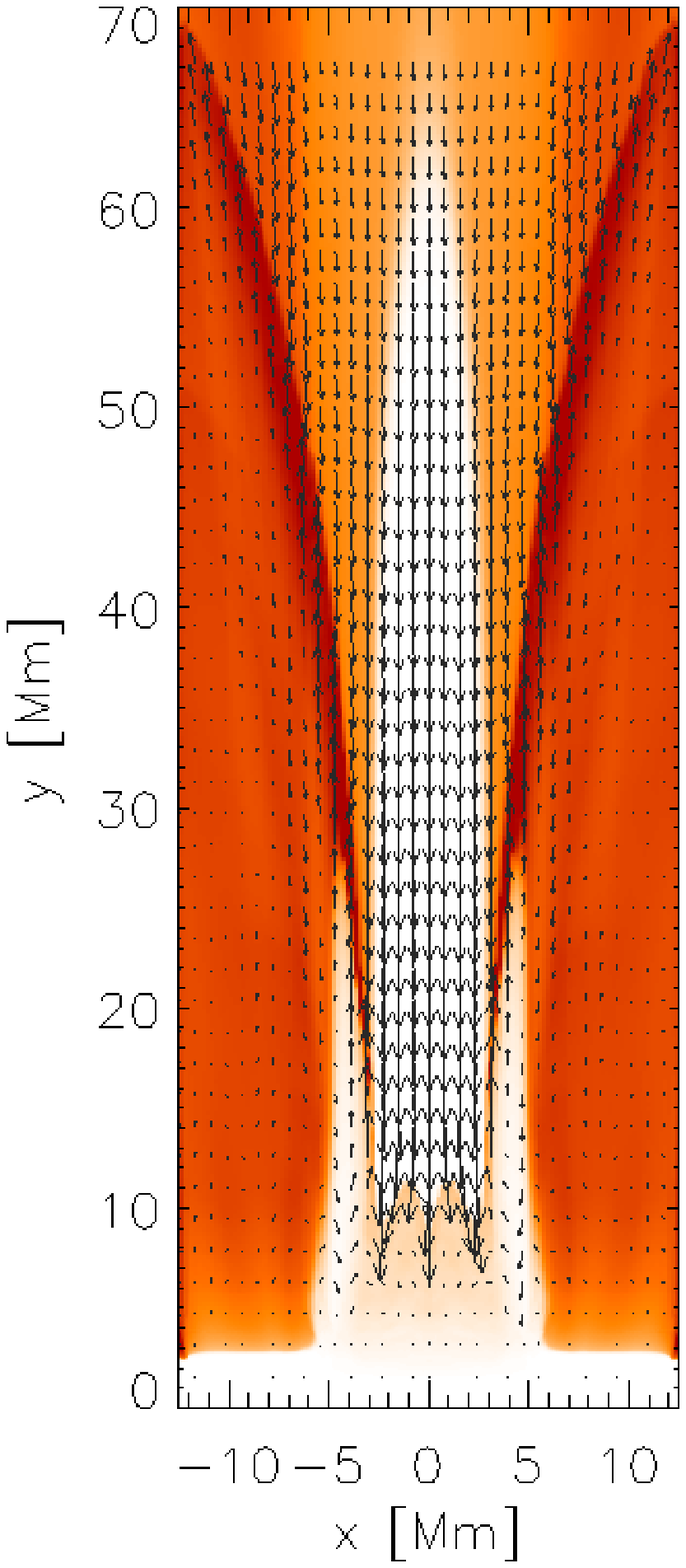}
  \includegraphics[scale=0.55]{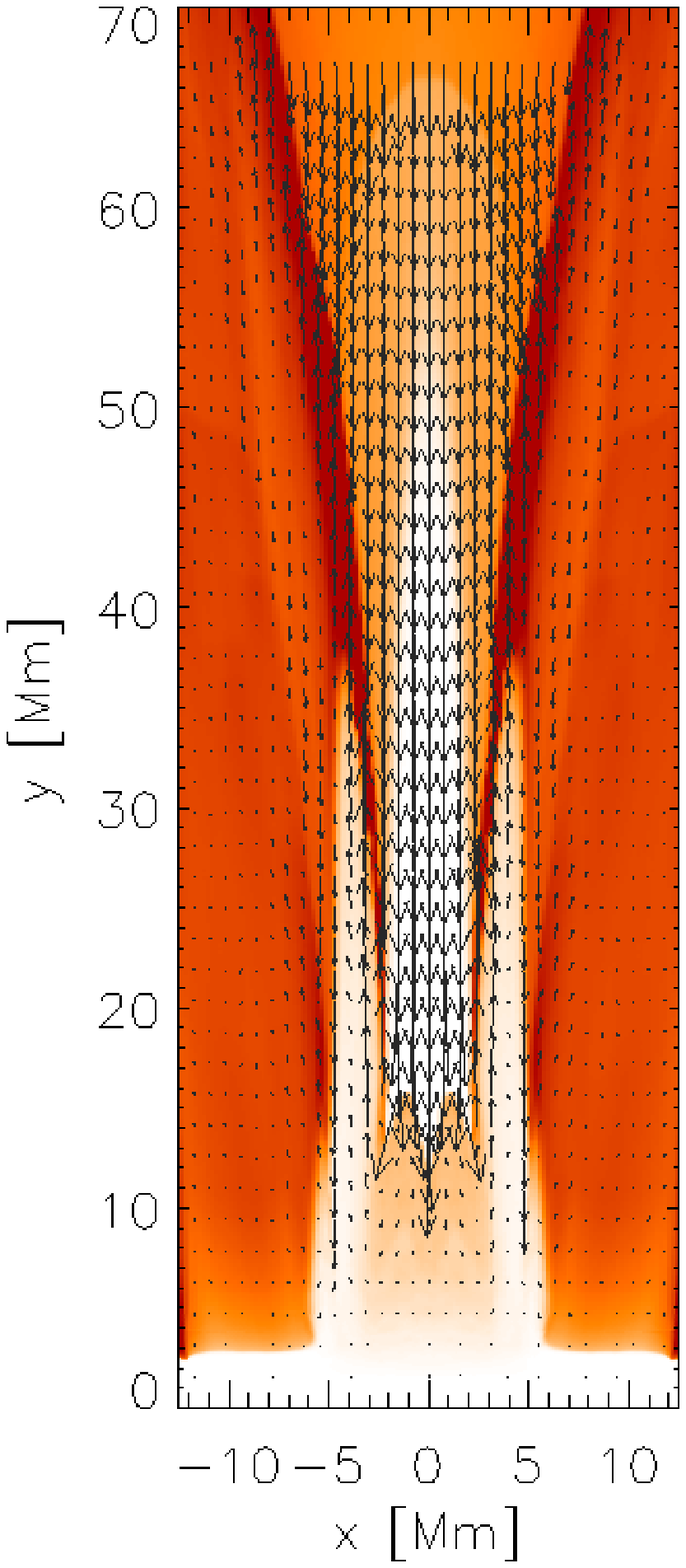}
  \includegraphics[scale=0.55]{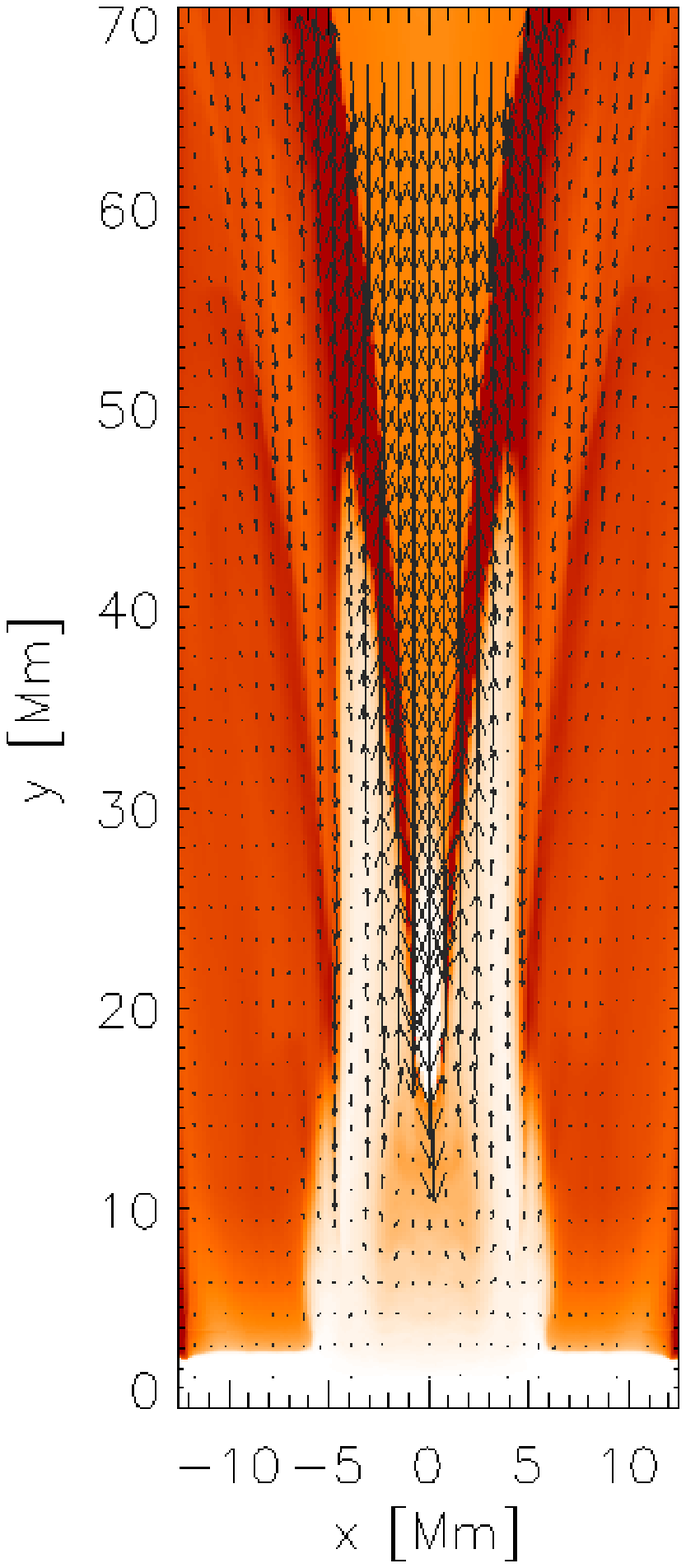}
 }
\caption{\small
Temporal snapshots of a simulated surge. 
The temperature (colour maps) 
profiles at $t=200$ s,
$t=400$ s, $t=600$ s, $t=800$ s, $t=1000$ s, and $t=1200$ s
(from top to bottom). 
Temperature is drawn in units of $1$ MK as shown in the colour bar, which is
common to all the panels. The velocity vector unit is 150 km s$^{-1}$.
}
\label{fig:serge_prof_cent}
\end{figure*}
%
\begin{figure*}

\centering
\mbox{
  \includegraphics[scale=0.55]{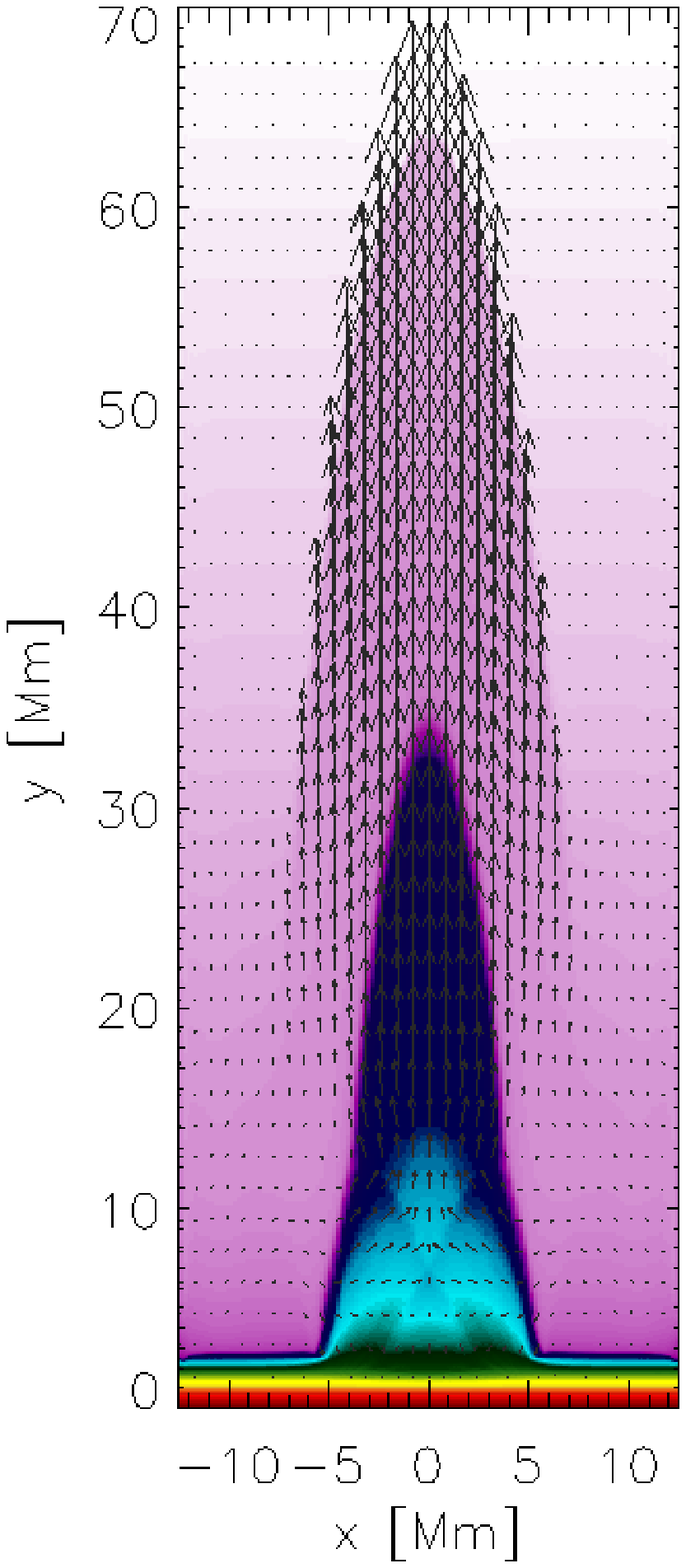}
  \includegraphics[scale=0.55]{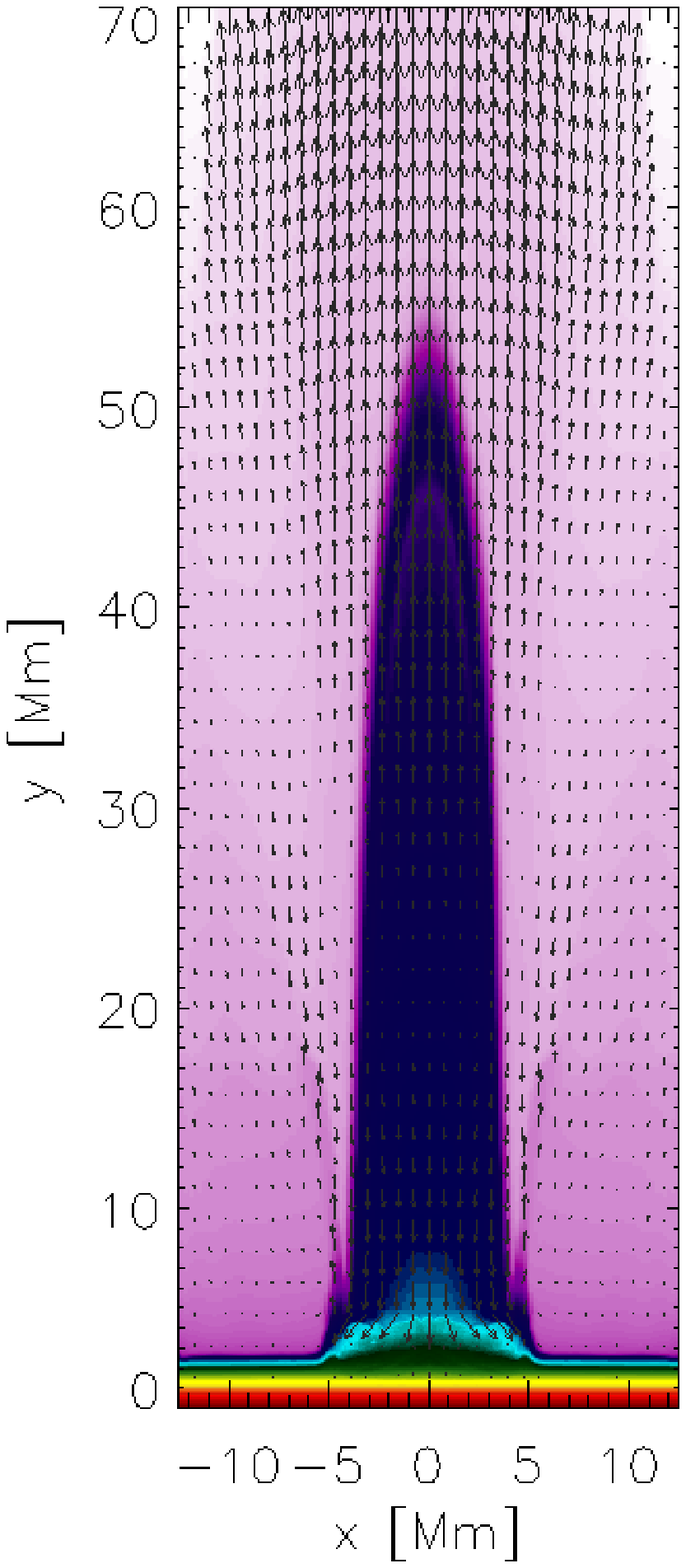}
  \includegraphics[scale=0.55]{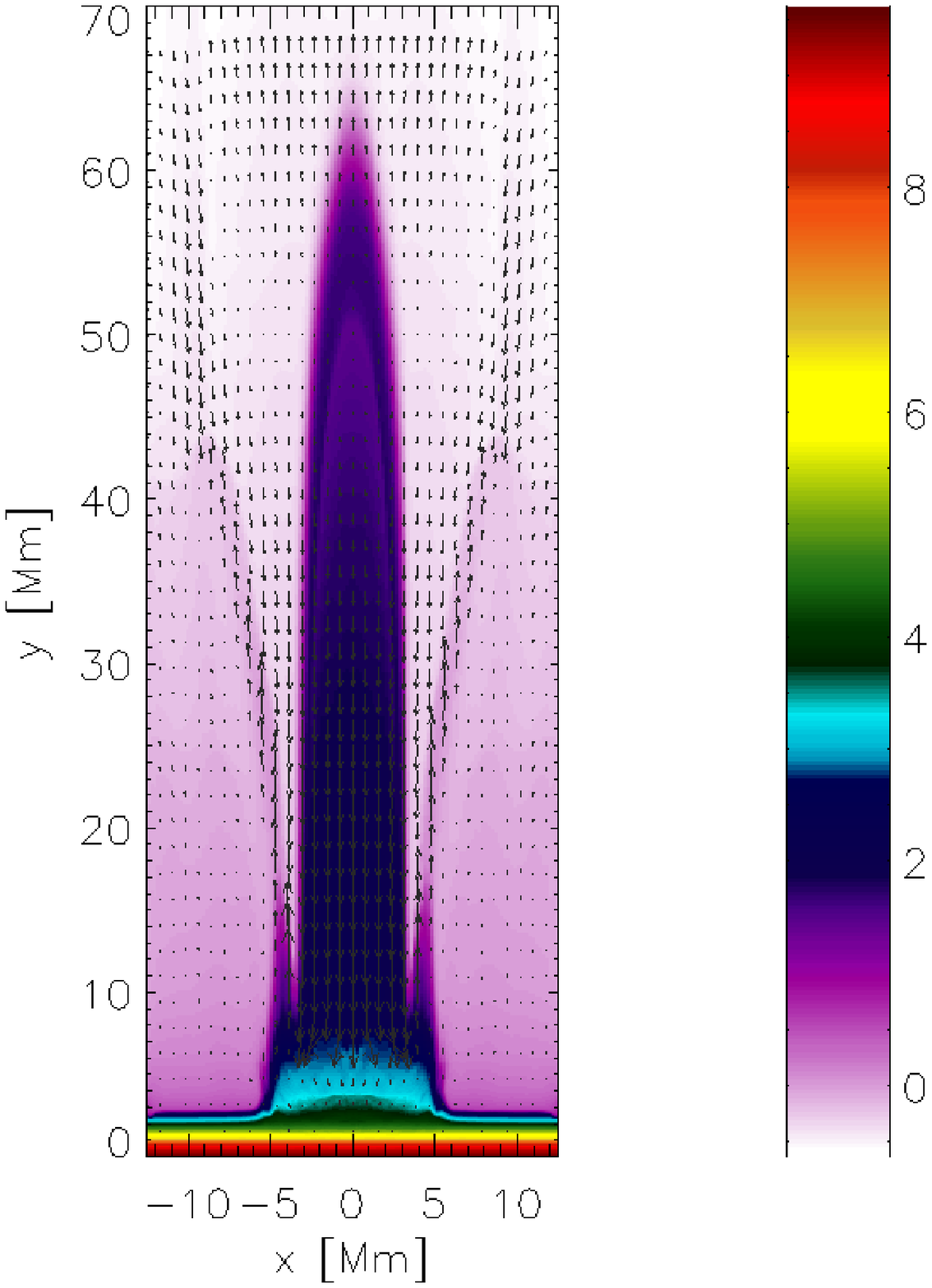}
}
\mbox{
  \hspace{-1.8cm}
  \includegraphics[scale = 0.55]{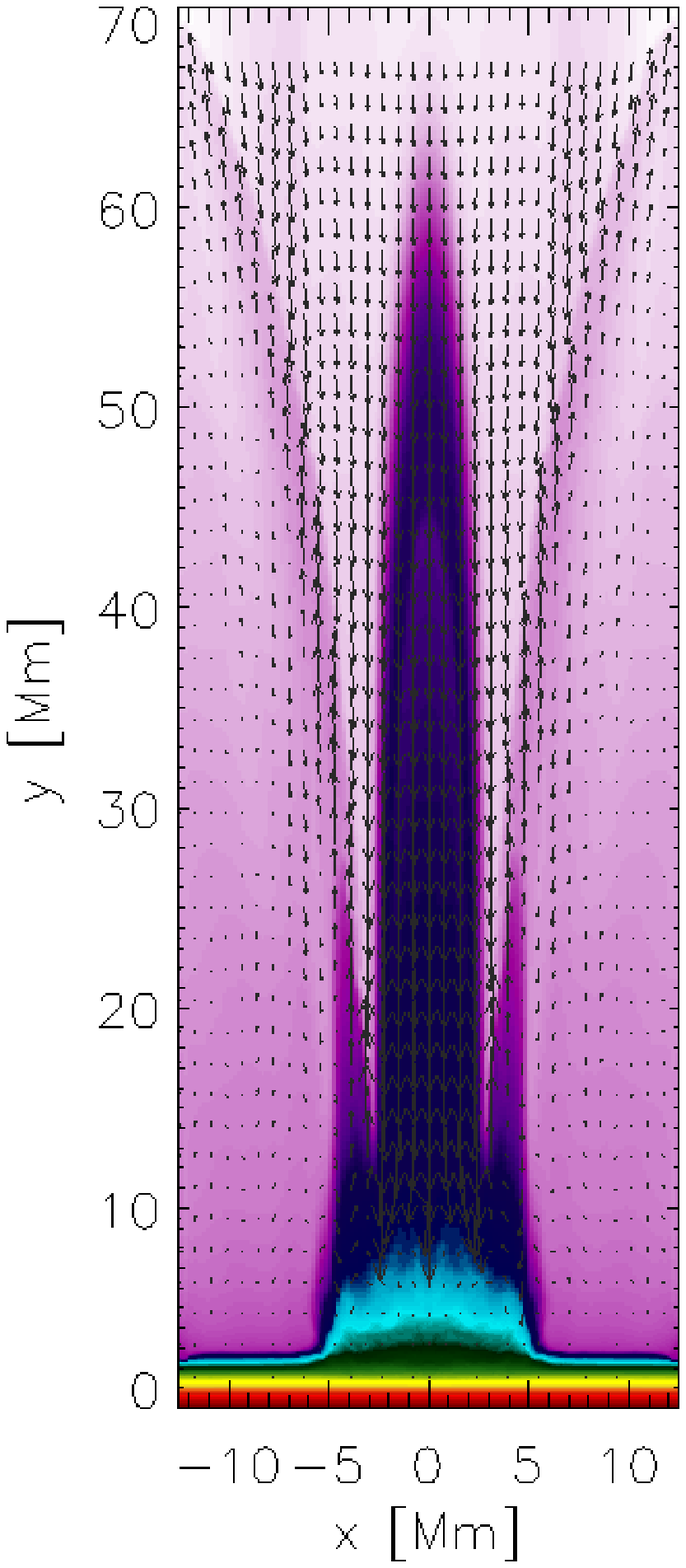}
  \includegraphics[scale=0.55]{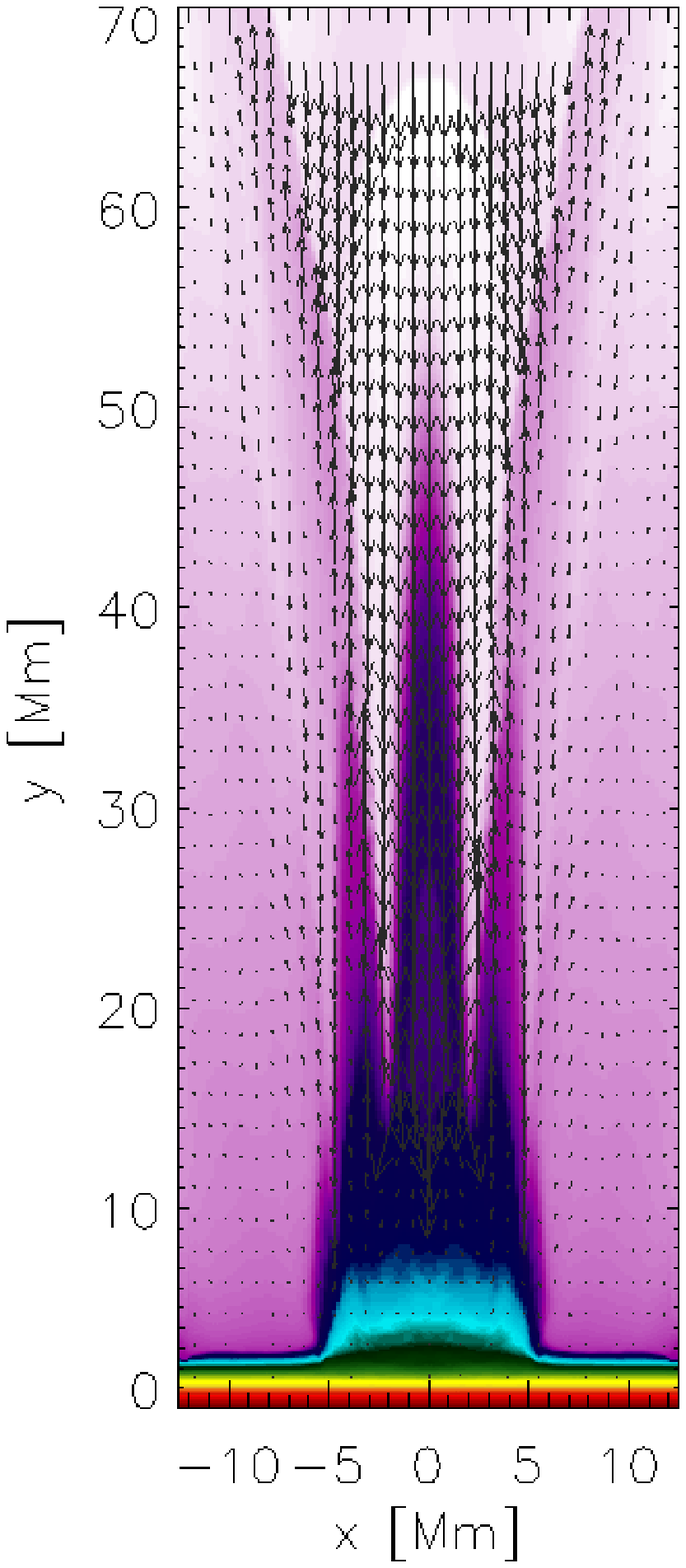}
  \includegraphics[scale=0.55]{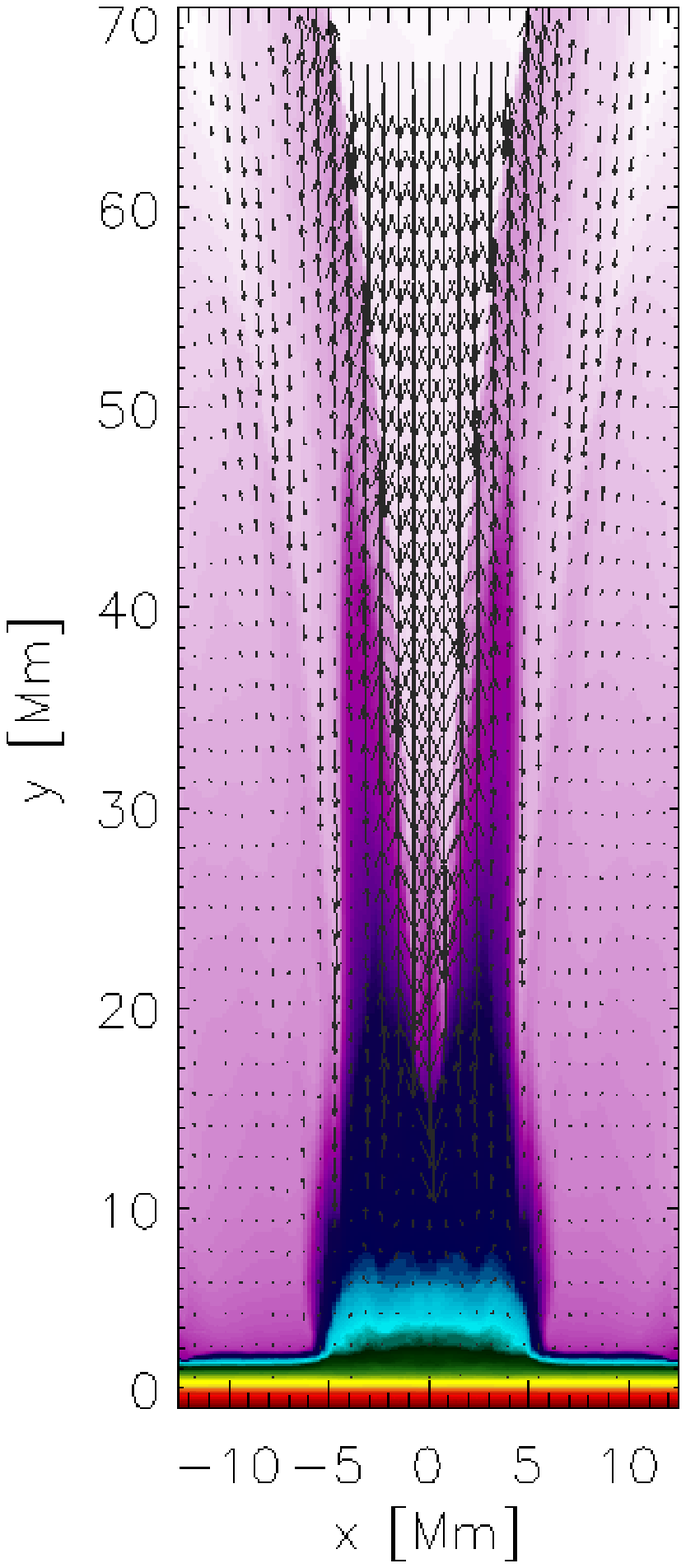}

}
\caption{\small
Temporal snapshots of a simulated surge. 
The density (colour maps) 
profiles at $t=200$ s,
$t=400$ s, $t=600$ s, $t=800$ s, $t=1000$ s, and $t=1200$ s
(from top to bottom). 
Density is drawn in the units of 10$^{-12}$ kg m$^{-3}$ as shown in the colour bar, which
is common to all the panels. The velocity vector unit is 150 km s$^{-1}$.
}
\label{fig:serge_prof_cent}
\end{figure*}






\clearpage

\end{document}